\documentclass[final,5p,times,twocolumn]{elsarticle} %% use this for final copy

\usepackage{xcolor} % Uncommented by Juan Abel Barrio (JAB) on 26/10/22 to allow for new text in color to identify different contributors

\usepackage[colorlinks=true]{hyperref}

\usepackage{epsfig}

\usepackage{amsmath}
\usepackage{amssymb}
 \usepackage{lineno}  % for submission only
\journal{Nuclear Instruments and Methods in Physics Research A}

\usepackage{adjustbox}
\usepackage{subcaption}
\usepackage{graphicx}

\newcommand{\labfig}[1]{\label{fig:#1}}
\newcommand{\reffig}[1]{\hyperref[fig:#1]{Figure}~\ref{fig:#1}\xspace}

\newcommand{\labsec}[1]{\label{sec:#1}}
\newcommand{\refsec}[1]{\hyperref[sec:#1]{Section}~\ref{sec:#1}\xspace}

\newcommand{\nc}[1]{\textcolor{black}{  #1}} % Corrections by nectarcam collaboration  (14/12/22) 

\newcommand{\sapoA}[1]{{\textcolor{black}{  #1}}} % Corrections by SAPO collaboration  (09/01/23) 
\newcommand{\sapoB}[1]{{\textcolor{black}{  #1}}}% Corrections by SAPO collaboration  (09/01/23) 

\newcommand{\nima}[1]{{#1}}% 
\newcommand{\nimabis}[1]{{#1}}% 
\usepackage{soul} % Added by JAB (26/10/22) for easy strike-out text

\usepackage{booktabs}

\begin{document}

\begin{frontmatter}



\title{The NectarCAM Timing System} %\\ [10pt]\large{(NectarCAM collaboration)}}

% Timing performance of the NectarCAM camera
%\def\thefootnote{\alpha{fnote}}

%\title{Elsevier Author List 20120813}
% \collab{NectarCAM Collaboration}
% \author{(NectarCAM Collaboration)}

\author[IRFU]{F.~Bradascio\fnref{corresponding1}}
\author[IRFU]{H.~Rueda}
\author[UCM]{J.A.~Barrio}
\author[IJCLab]{J.~Biteau}
\author[IRFU]{F.~Brun}
\author[APC]{C.~Champion}
\author[IRFU]{J-F.~Glicenstein}
\author[CPPM]{D.~Hoffmann}
\author[TOU]{P.~Jean}
\author[LPNHE]{J.P.~Lenain}
\author[IRFU]{F.~Louis}
\author[UCM]{A.~P\'erez}
\author[APC]{M.~Punch}
\author[IRFU]{P.~Sizun}
\author[DESY]{K.-H.~Sulanke}
\author[UCM]{L.~A.~Tejedor}
\author[IRFU]{B.~Vallage}

% \author[]{+ others}
% % \author[MadisonPAC,MadisonPAC]{F.~Brun\fnref{corresponding1}}
% % \author[MadisonPAC,MadisonPAC]{S.~Miarecki\fnref{corresponding2}}

\fntext[IRFU]{IRFU, CEA, Universit\'e Paris-Saclay, F-91191 Gif-sur-Yvette, France}
\fntext[CPPM]{Aix-Marseille Université, CNRS/IN2P3, CPPM, 163 Avenue de Luminy, 13288 Marseille cedex 09, France}
% Updated MP on 2022-12-10:
% Modifying the address at : https://u-paris.fr/charte-des-signatures-de-publications-scientifiques-universite-de-paris/ adding IN2P3 and APC acronym
\fntext[APC]{Universit\'e Paris Cit\'e, CNRS/IN2P3, AstroParticule et Cosmologie (APC), Paris~F-75013, France}
\fntext[UCM]{IPARCOS-UCM, Instituto de Física de Partículas y del Cosmos, and EMFTEL Department, Universidad Complutense de Madrid, E-28040 Madrid, Spain} % Uddated by JAB on 26/10/22
\fntext[IJCLab]{Université Paris-Saclay, CNRS/IN2P3, IJCLab, 91405 Orsay, France}
\fntext[TOU]{Institut de Recherche en Astrophysique et Plan\'etologie, CNRS-INSU, Universit\'e Paul Sabatier, 9 avenue Colonel Roche, BP 44346, 31028 Toulouse Cedex 4, France}
\fntext[LPNHE]{Sorbonne Université, Université Paris Diderot, Sorbonne Paris Cité, CNRS/IN2P3, Laboratoire de Physique Nucléaire et de Hautes Energies,
LPNHE, 4 Place Jussieu, F-75252 Paris, France}
\fntext[DESY]{DESY, D-15738 Zeuthen, Germany}
\fntext[corresponding1]{Corresponding author:  Federica Bradascio, Email \href{mailto:federica.bradascio@cea.fr}{federica.bradascio@cea.fr}}
% \fntext[corresponding2]{Corresponding author:  Spencer Klein, Lawrence Berkeley National Laboratory, 1 Cyclotron Rd, Mail Stop 50R5008, Berkeley CA 94720 USA, Phone +1-510-486-5470, Fax +1-510-486-6738, Email srklein@lbl.gov } 

\begin{abstract}
% The uncertainty on the trigger timestamp of a Cherenkov camera relative to the time of arrival of light at the detection plane and the estimation of the light's arrival time in each pixel are important information that can be used to reduce the noise in shower images and improve the imaging cleaning and discrimination between Cherenkov photons and background. In this paper, we present results on the timing performances of NectarCAM.NectarCAM is a Cherenkov camera which is going to equip the medium-sized telescopes (MST) of the north site of the Cherenkov Telescope Array Observatory (CTAO). NectarCAM is equipped with 265 modules, each consisting of 7 photo-multiplier tubes (PMTs) and a Front-End Board performing the data capture. The tests have been performed on the first NectarCAM unit in the dark room of the {integration and test facility} in CEA Paris-Saclay (France). 

NectarCAM is a Cherenkov camera which is going to equip the \sapoB{Medium-Sized Telescopes} (MST) of the northern site of the Cherenkov Telescope Array Observatory (CTAO). NectarCAM is equipped with 265 modules, each consisting of 7 photo-multiplier tubes (PMTs), a Front-End Board and a local camera  trigger system used for data acquisition.
This paper addresses the timing \nima{performance} of NectarCAM which are crucial to reduce the noise in shower images and \sapoB{improve image cleaning} as well as {to} discriminate between gamma-ray photons and cosmic-ray background {and finally to allow coincidence identification with neighbouring telescopes for stereoscopic operations}.
\sapoB{Verification tests of the system} have been performed in a dark room using various light sources to illuminate the first NectarCAM unit. \sapoA{The resulting timing precision and accuracy of the trigger arrival relative to a laser source, of individual and multiple pixel signals have been studied and are shown to comply to CTAO requirements.}
% \sapoB{The resulting timing precision and accuracy} of the trigger arrival, of individual and multiple pixel signals have been studied and are shown to comply to CTAO requirements.
% The timing accuracy of the trigger arrival, of individual and multiple pixel signals has been studied and were shown to comply to CTAO requirements.
\end{abstract}


\end{frontmatter}

 % \linenumbers  %% for submission only

\begin{figure*}[th]
    \centering
    \includegraphics[width=0.9\textwidth]{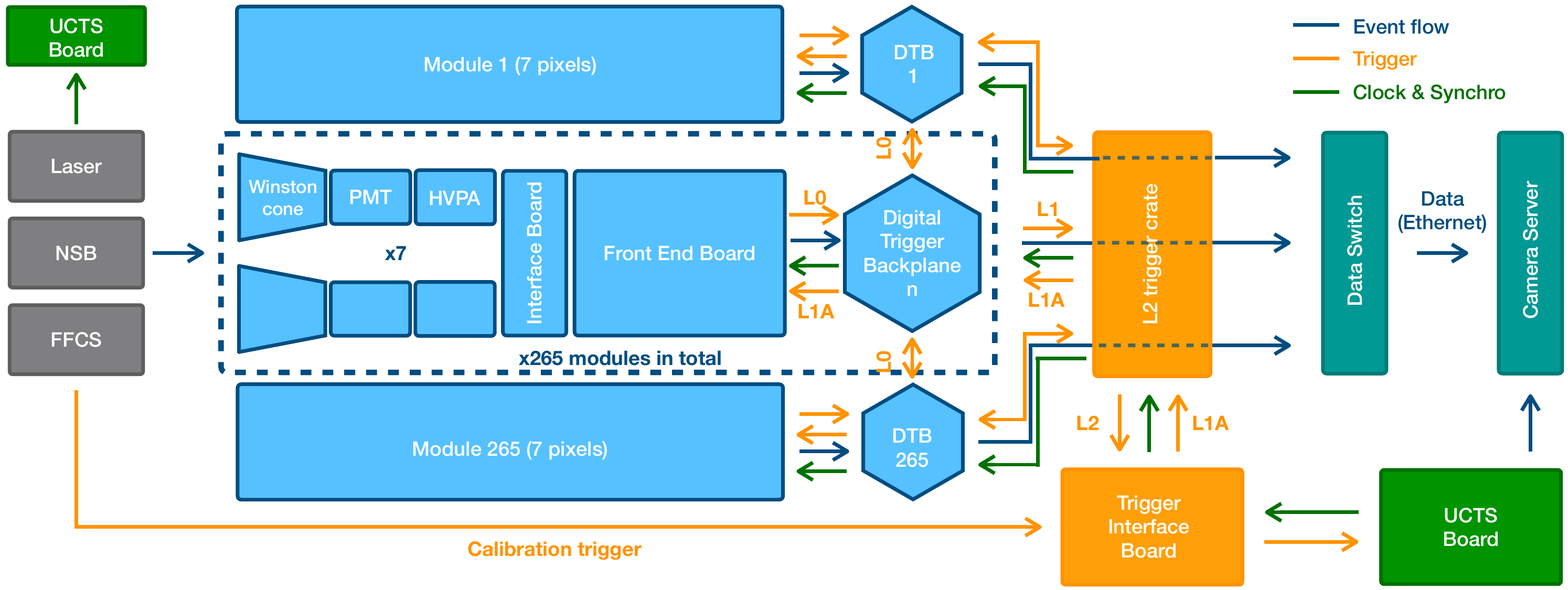}
    \caption{\sapoB{Schematic illustration of the signal and trigger chain of NectarCAM. The main detection body of the camera is shown in blue. The camera server is located outside the camera frame, in the CTAO on-site data center farm. The light sources used in the dark room for the verifications described in this paper are also shown on the left in grey. The laser source is connected to a \nima{Unified Clock distribution and Trigger time-Stamping (UCTS)} board that records the timestamp of the photons sent to the camera.}}
    \labfig{trigger}
\end{figure*}

% The L0 signals for a cluster of 7 PMT pixels are generated on the front-end board (FEB) and sent to a cluster FPGA on the Digital Trigger Backplane (DTB) board. Each cluster FPGA exchanges L0 signals with the FPGAs assigned to neighbouring clusters (L0 neighbour) and executes a trigger algorithm on 37 channels to generate the L1 signal for its super-cluster. The design for the L2 stage foresees 18 Cluster Service Boards (CSB) that accept L1 signals from 16 super- clusters; the total number of pixels is then 2016. The camera trigger is finally generated by one L2 Controller Board (L2CB) based on the input from the 18 CSBs.

\section{Introduction}
% \begin{itemize}
%     \item Introduction on CTA, MST and NectarCAM
%     \item  Motivation for timing calibration
% \end{itemize}

The imaging atmospheric Cherenkov technique is \nima{one of the major methods} for observations of very-high energy gamma rays on Earth, \nima{the other being the water Cherenkov technique \cite{hawc,lhaaso}.} In the imaging atmospheric Cherenkov technique, gamma rays are detected indirectly via the Cherenkov light produced in particle cascades (showers) generated when they interact in the atmosphere.
% In this technique, gamma rays are detected indirectly via the Cherenkov light emitted by the charged particles of the induced air-shower.
Images of the shower in one or several telescopes are analyzed to provide \sapoB{cosmic-ray} background suppression and to reconstruct primary \sapoB{gamma-ray} parameters.

\nc{CTAO} is {the major future facility} in the field of astroparticle physics and high-energy astrophysics exploiting the imaging atmospheric Cherenkov technique~\cite{CTAconcept}. The CTA \sapoB{Observatory} \nima{will explore} the high-energy universe with gamma rays between 20~GeV {and} 300~TeV~\cite{cta_mc_design}. It \sapoA{consists} of {several tens of} telescopes distributed over two sites, in La Palma (Spain) and in Paranal (Chile), in three different sizes: Small-Sized Telescope (SST), Medium-Sized Telescope (MST) and Large-Sized Telescope (LST). The telescopes consist of tessellated mirrors which focus the Cherenkov photons arriving within a few nanoseconds onto fast-recording, pixelated cameras. In order to properly combine the spatial and temporal information of these short light pulses from all the telescope cameras and accurately reconstruct the properties of the observed air shower, precise knowledge of the shower image timing is mandatory.

In this paper, we present the timing \nima{performance} of NectarCAM, the camera designed to equip the MST telescopes at the CTA-North site~\cite{nectarcam}. \nima{The Cherenkov signal arrives in individual pixels in a time interval of 2--4~ns. The relative timing of individual pixels can be used to reduce the night sky background for low-energy gamma-ray showers, which give signals in just a few pixels in the camera. On the other hand, high-energy showers can give images with a time gradient in the camera that can last up to several tens of ns. The relative timing information between pixels can be used to calculate the shower time gradient. Finally, stereoscopy allows to reduce the camera trigger and acquisition rate by rejecting events that are not seen by several telescopes.  For this, the relative time between telescopes must be known with an accuracy of tens of ns, but in addition the geometry of the Cherenkov wavefront can be measured if an accuracy of a few ns is achieved.
Therefore, there are three types of timing uncertainty that need to be characterized:} the single pixel precision, limited by the \nima{quantization} noise of the chip responsible for the signal read-out; the total camera timing precision, which takes into account the timing difference between each \nc{pair} of pixels, and the timing accuracy associated to the camera trigger timestamp.
Each of these timing uncertainties has been quantified with the first NectarCAM camera unit in the darkroom at the {integration and test facility} \nima{at CEA (Commissariat à l'Énergie Atomique et aux Énergies Alternatives)} Paris-Saclay.
These tests are also required by CTAO since it follows a verification-based product acceptance procedure, according to which all products must fulfill a list of performance requirements before their deployment on site. 
% \nima{A list of the requirements tested in this paper is shown in \reftab{ctao_requirements}.}

% These tests are also required by CTAO before the deployment of the camera on site. CTAO follows a verification-based product acceptance procedure, according to which all products must fulfill a list of performance requirements. 

% In the following,
% The {timing} tests that have been performed with the first NectarCAM camera unit in the darkroom at the {integration and test facility} in CEA Paris-Saclay (France).

% Tis is also required by the CTAO compliant with the CTAO acceptance procedures. CTAO follows a verification-based product acceptance procedure

% We present performance results of the NectarCAM qualification model, which focus on the characterization of the timing accuracy and systematic uncertainties. The tests have been performed with the first fully equipped NectarCAM camera (265 modules) .
% Results on the timing accuracy and on the systematic uncertainties for each pixel and the full camera, and the performance of the camera trigger are described.
% {In the following, the results on the timing accuracy and systematic uncertainties for each single pixel and for the full camera are presented together with the performance of the camera trigger. }
\sapoB{In the following, the timing accuracy and systematic uncertainties are presented for both single pixels and the full camera together with the overall performance of the camera trigger.}

% \begin{table}[]
%     \centering
%     \resizebox{\textwidth}{!}{%
%     \begin{tabular}{c | c | c}
%         \toprule
%          \sc Requirement & \sc Scientific Motivation & \sc Value\\
%          \midrule
%          \midrule
%          Single pixel timing precision & The time dispersion for an instantaneous light pulse  must be shorter than the intrinsic Cherenkov pulse duration of 4~ns per pixel. & $<1$~ns \\
         
%          Global camera timing precision & The relative arrival time of light in different camera pixels is desirable for the noise reduction in images to improve shower reconstruction.  & $<2$~ns \\
         
%          Camera trigger timing accuracy & For event synchronisation purposes, the inter-telescope trigger timing accuracy must be $<10$~ns per event. & $<2$~ns\\
%          \bottomrule
%     \end{tabular}}
%     \caption{The requirements tested in this work with the corresponding scientific motivation are listed. The official requirement are not public.}
%     \labtab{ctao_requirements}
% \end{table}

\section{NectarCAM}
% Detailed and technical description of the camera and of the trigger system
\sapoB{NectarCAM} has a modular design, with a basic element (``module") consisting of a focal plane module (FPM) and a front-end board (FEB). The \sapoB{FPM} \cite{2021NIMPA100765413T} is composed of seven 1.5" R12992-100-05 Hamamatsu photomultiplier tubes (PMTs) associated with high voltage and pre-amplification boards (HVPA) \sapoA{sitting in the FEB}, and equipped with Winston cone light concentrators \cite{lightconcentrators}. 

\subsection{Camera readout}
\sapoB{Cherenkov light impinging on the camera is converted into an \nima{electrical} signal} by the PMTs and preamplified into two (high and low {gain}) signals.
\nc{The \sapoA{Amplifier for CTA (ACTA)~\cite{acta}} \sapoB{application specific integrated circuit (ASIC)}} chip of the FEB amplifies again the high gain signal and {splits} the output into \nima{the} high gain signal input \nima{to} the NECTAr chip~\cite{nectar0} and \nima{the} trigger level 0 (L0), while the low gain \nima{signal} is simply amplified to enter another channel of the NECTAr chip.
% The signal is then amplified a second time in the FEB by a custom amplifier called ACTA, where it is split into three channels: low and high gain channels, and trigger channel (level 0 or L0 trigger). 
{The latter comprises} a switched capacitor array \nima{which samples} the signal at 1~GHz\sapoB{,} {and} a 12-bit analogue-to-digital converter (ADC), responsible for the digitization of each sample \sapoA{once the trigger signal is received}.
% The sampling and digitization of the signal is performed in the NECTAr chip~\cite{nectar0}, a switched capacitor array able to perform the sampling of the signal at 1 GHz, combined with a 12-bit analogue-to-digital converter (ADC). 
\sapoB{The NECTAr chip acts as a circular} buffer which holds 1 microsecond of data until a camera trigger occurs.
The camera is triggered in a multistep process (see the following section for details). When the trigger conditions are met, the photon signal is digitized, and  the full waveform for every pixel sent to a \sapoB{camera server} through an ethernet-based data acquisition system. % records the full waveform for every pixel. 
The camera server is a remote machine located outside the camera frame, in the CTAO on-site data center farm running the NectarCAM acquisition software. The data are transferred via 10~Gbps optical fibres. An absolute time-stamp is also assigned to the trigger signal (see \ref{sec:timing} below), so that images from multiple telescopes can be later combined.

\subsection{Camera trigger}
%  (of the order 0.2 deg$^2$, corresponding to circa three pixels) 
{NectarCAM} is triggered when it receives a significant amount of light in a compact region of the detection plane within a short period of time (typically a few ns)~\cite{trigger}. Translated in terms of camera hardware, this means that several neighboring pixels will see photon signals within a few nanoseconds. A schematic of the hardware architecture of the digital trigger of NectarCAM is shown in \reffig{trigger}.
%The final event trigger signal is {distributed by} the trigger interface board (TIB)~\cite{TIB}, after gathering the different trigger sources.
% The basic strategy is thus to trigger first a neighbouring group of pixels (level 0 trigger or L0) by the L0 application specific integrated circuit (ASIC) in the FEB, and then the camera (level 1 or L1, and level 2 or L2) by combining the digitized L0 signals at the digital trigger (DT) system. The final event trigger signal is formed at the trigger interface board (TIB)~\cite{TIB}, after gathering the different trigger sources.
\begin{figure}[h!]
    \centering
    \includegraphics[width=0.7\columnwidth]{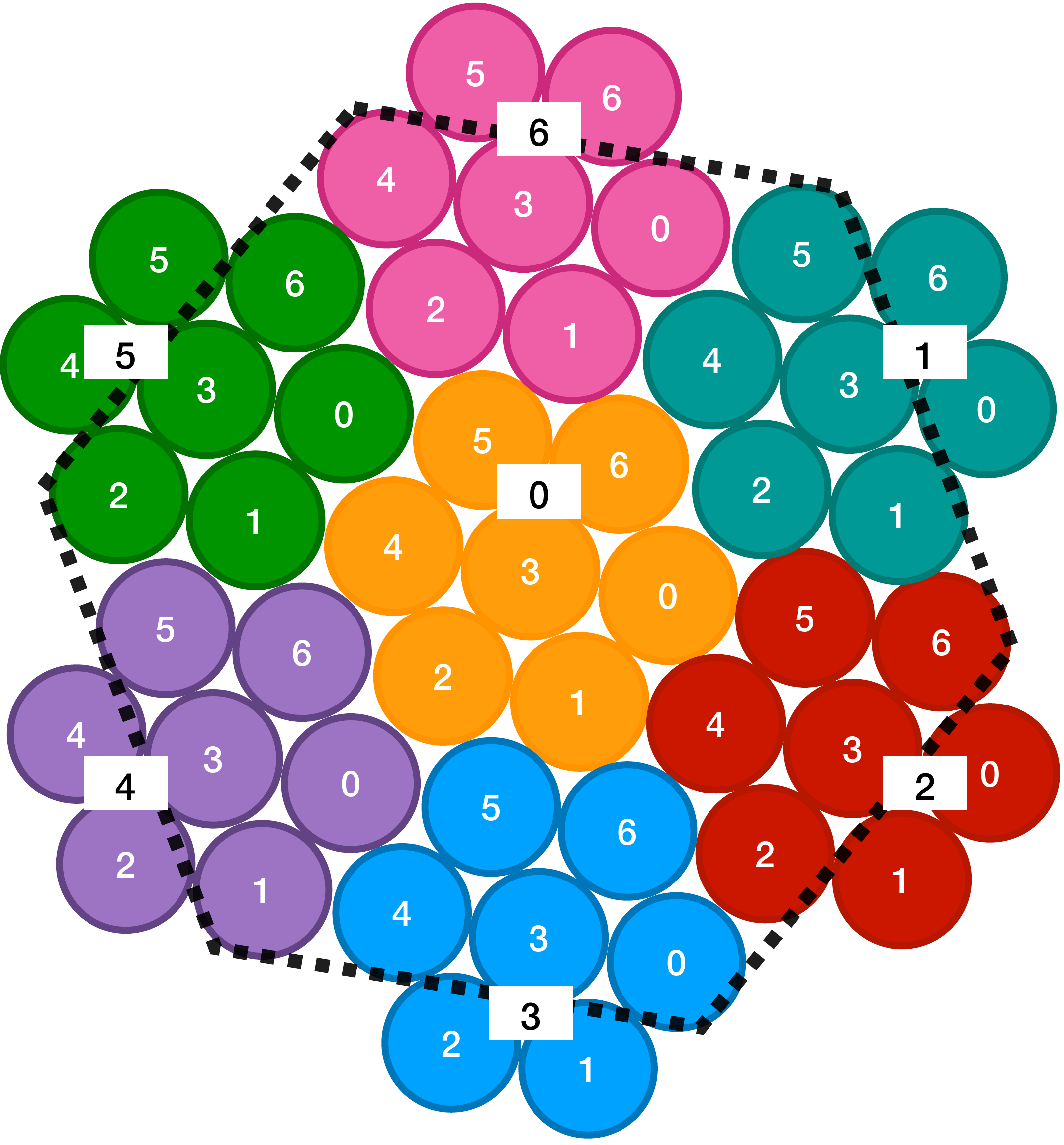}
    \caption{\nc{Schematic illustration of the 37 neighboring pixels where the L1 trigger is formed by combing the L0 signals: seven signals are formed in the central cluster (marked with 0) and five signals in each of the six surrounding clusters (cluster 1 to 6). \sapoA{Each circle represents one PMT, numbered from 0 to 6. PMTs with the same color are part of the same module. The dashed black line encloses the 37 pixels used to form the L1 trigger.}}}
    \labfig{37pixel_region}
\end{figure}
The first trigger level (level 0 trigger or L0)~\cite{L0} occurs in each FEB \nc{in \sapoB{an ASIC}: it} processes the analogue signals coming from the seven individual PMTs in each module, by comparing each pixel amplitude with a threshold value. \sapoB{If the signal in a pixel exceeds the settable threshold, a digital L0 signal is produced for that pixel.} 
% If it is above it, a L0 digital signal is associated to the corresponding pixel.
% Then, the DTB of a given module receives the L0 signals from the pixels of that module plus the L0 signals from most of the pixels in surrounding modules, as shown in Figure 2, and produces its corresponding L1. Therefore, L0 is pixel-wise, L1 is camera-region wise and L2 is full-camera-wise.   
The L0 digital signal is then sent to an FPGA located on the digital trigger backplane (DTB) of that module, where the level~1 (L1) trigger signal is formed by processing the L0 signals of that module and signals from {the nearest 5 of the 7 PMTs in} the 6 neighboring modules, forming a 37-pixel region, \sapoB{as indicated by the area contained within the dashed black line in \reffig{37pixel_region}}.  
% Each module's DTB forms an L1 signal \sapoB{from its} neighboring L0 signals, therefore ensuring trigger homogeneity throughout the camera. The L1 trigger is created by the DTB's FPGA using the 3 nearest neighbours (3NN) algorithm~\cite{trigger}. 
\sapoA{Each module's DTB uses the 3 nearest neighbours (3NN) algorithm~\cite{trigger} implemented in its FPGA to form a L1 trigger signal out of the 37 L0 signals.} \sapoB{The 3NN algorithm was tested with simulations and real data~\cite{3nn_mc}.}

The FPGA distributes the L0 signals of a module to its six neighbours and in parallel receives their L0 outputs (see \reffig{37pixel_region}). \sapoB{The delay of each of the 37 L0 signals can be adjusted digitally in steps of 50 ps, before entering the trigger chain.}
% Each of the 37 L0 signals has to pass its (FPGA-internal) individual {digital} delay line, before entering the trigger chain. The individual delays can be adjusted in steps of 50 ps. 
\nc{The 3NN trigger algorithm} generates a trigger, if within the 37 pixel area 3 adjacent pixels within a 3~ns {discrete} time window are above a discrimination threshold.  
The time window width of 3~ns is the default value on the first NectarCAM. This is justified by the study illustrated in \reffig{trigger_window}, which shows the rate of occurring L1 triggers \sapoB{within three different} time windows. The delay on the \nima{horizontal axis} has been obtained by directly injecting light into \sapoB{three} connected pixels (either oriented inline or triangular) belonging to \sapoB{three} neighboring modules with optical fibers. \nima{The optical fibers produce a short duration pulse approximately 5~ns wide (full-width half-maximum,  FWHM), with around 3~ns rise time (from 10\% to 90\% of full amplitude) \cite{KAPUSTINSKY1985612}.} The delay of one fiber with respect to the others has been changed {and three different values of time window (1~ns, 3~ns and 5~ns) have been tried.} \nima{As} can be seen in \reffig{trigger_window}, {the 1~ns time-window misses more than 40\% of the triggers. The 3~ns time-window is \nima{more than 95\%} efficient \nc{in the interval -1~ns to 0~ns} and integrates less noise than the 5~ns time-window.} 
% While injecting 3 signals into 3 channels of module 244, we measured the TIB local camera rate as we delayed 1 signal w.r.t. the others.

\begin{figure}
    \centering
    \includegraphics[width=\columnwidth]{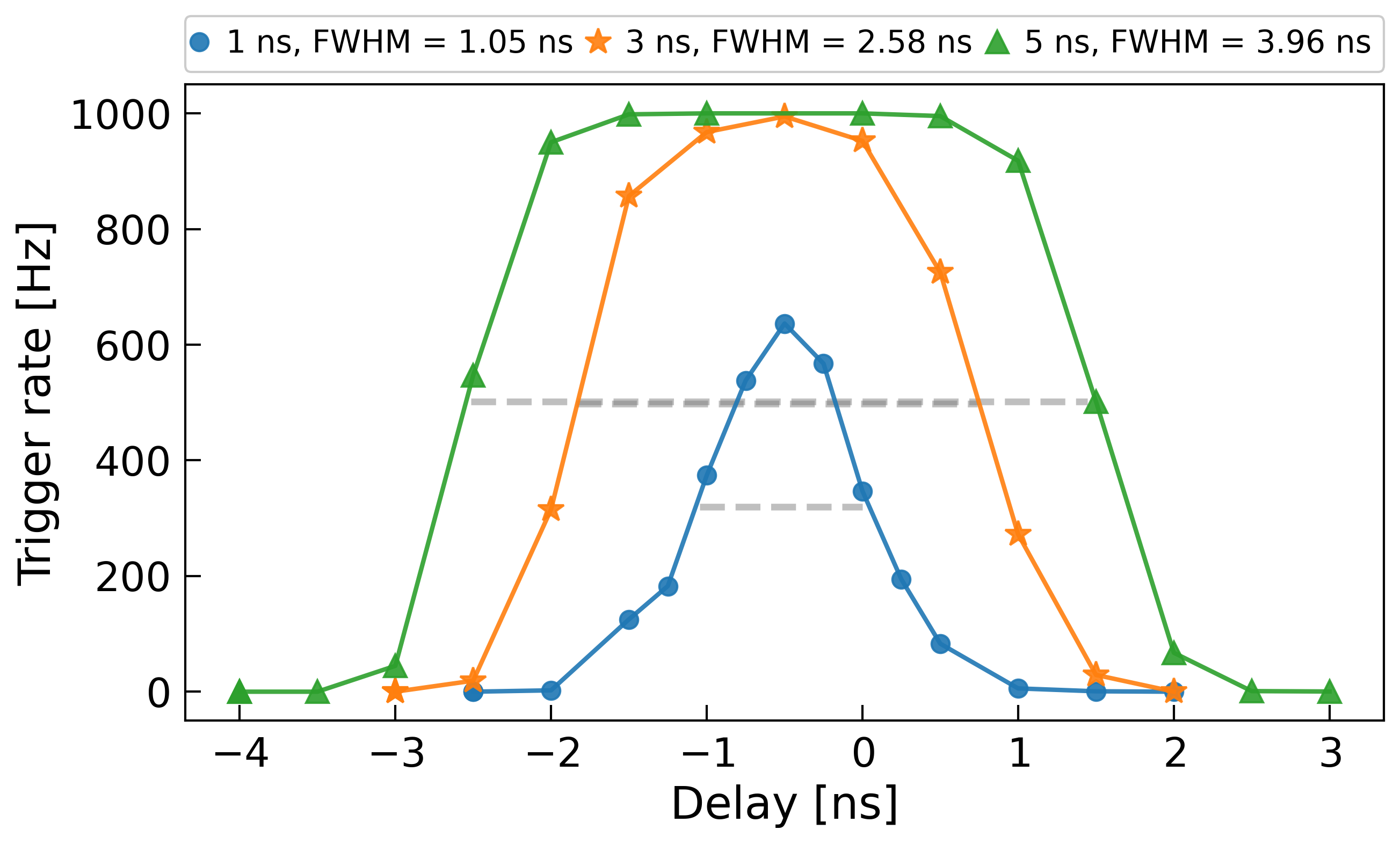}    
    \caption{Trigger rate as a function of the time delay between 3 optical fibers injecting light directly into 3 pixels in 3 neighboring modules to test the 3NN trigger time window. \nima{An arbitrary common delay of 6.5~ns was added to all three channels resulting in an arbitrary position of the 0~ns delay.}}
    \labfig{trigger_window}
\end{figure}

{The L1 signals of each DTB (265 in total) are delivered to an L2-crate (L2C) where they are combined in an OR operation to generate the L2 trigger, i.e., the camera trigger, which is then sent to the trigger interface board (TIB)~\cite{TIB}.}
% The L1 signals of each DTB (265 in total) are delivered to a clock and trigger distribution board (CTDB). There are 18 CTDBs, each serving up to 15 DTBs, which are plugged into a L2-crate (L2C) together with the L2 controller board (L2CB). The L2CB processes the L1 signals from all the camera modules and combines them into a OR operation to generate the L2 trigger, i.e., the camera trigger, which is thus sent to the TIB.
% \nc{\st{Once \jab{\st{the event} this camera} trigger signal is formed at the TIB, it is \nc{\st{delivered back} returned} as a L1-accepted (L1A) signal all the way back to the FEBs.}
{Depending on the configuration parameters and camera status, the TIB ignores the L2 or delivers it back as a L1 accepted (L1A) signal all the way back to the FEBs.}
{At this point}, the 1~GHz~sampling of the NECTAr chips is stopped, {the full 60~ns waveform} for every pixel is digitized and transferred over ethernet to a \sapoB{camera server} \nc{where the different components of the camera event are combined in an Event Builder process.} In addition, the TIB delivers the {camera} trigger signal and the corresponding \sapoA{trigger-class bit associated to a type of event (e.g. pedestal, calibration, internal)} to the Unified Clock distribution and Trigger time-Stamping (UCTS) module, where the event is time-stamped, as described below. 

% In addition, the TIB delivers the event trigger signal to the unified clock distribution and trigger time-stamping (UCTS) board, where the event is time-stamped on the TiCkS module~\cite{ticks}, a dedicated White Rabbit based board. 
\begin{figure}
    \centering
    \includegraphics[width=\columnwidth]{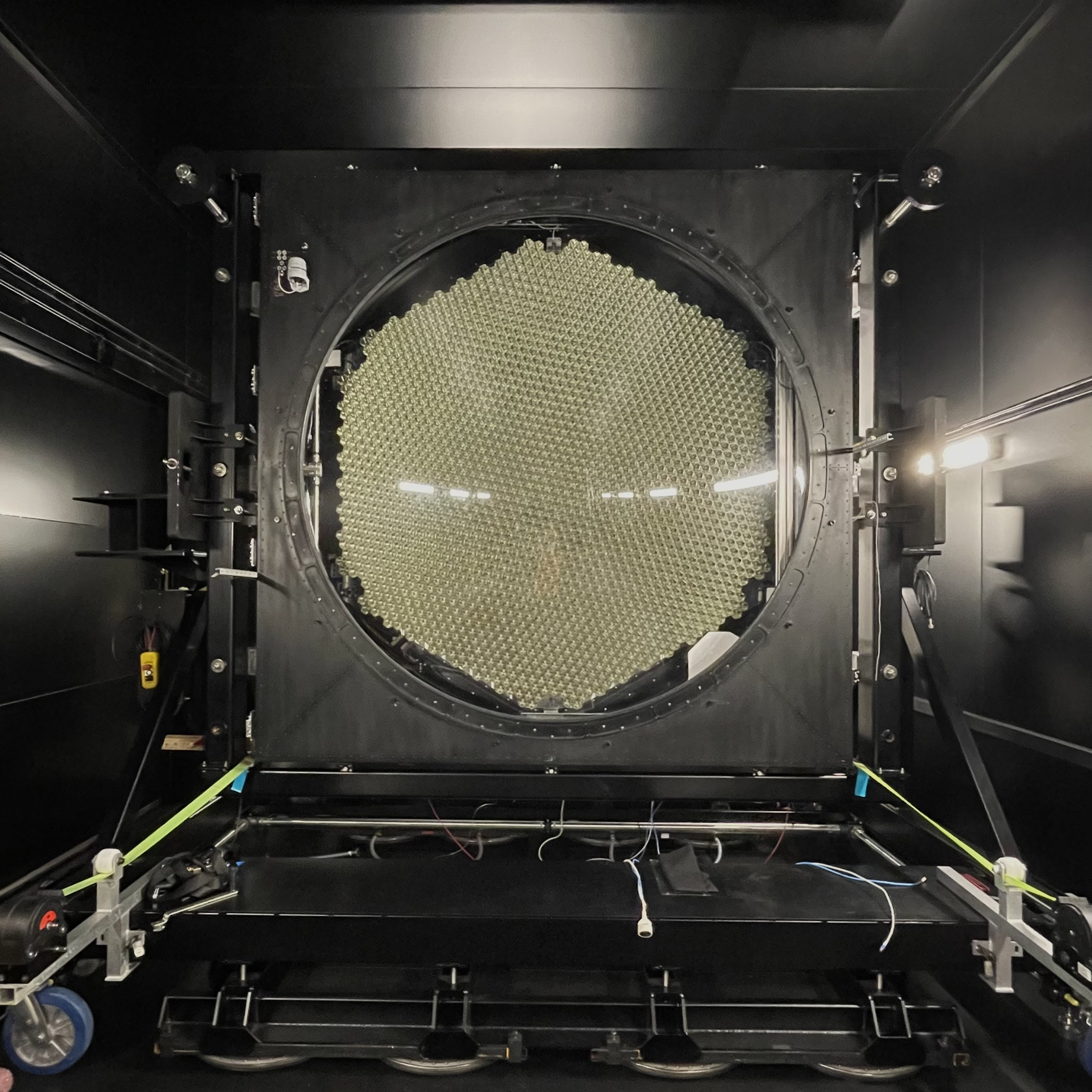}
    \caption{NectarCAM camera with entrance window in the dark room of CEA Paris-Saclay. {It has a size of $2.9\times 2.8\times1.5$~m and weighs about two tons.} The full camera is equipped with 1855 PMTs. \nima{The white target used for single photoelectron calibration is partially visible in the lower right corner of the camera \cite{xy_table}. This system has not been used for any calibration described in this paper.}}
    \labfig{nectarcam}
\end{figure}

\subsection{Camera absolute time-stamping system}
\label{sec:timing} 

The UCTS module contains primarily a timing and clock stamping (TiCkS) board~\cite{ticks}, a node of \nc{a} White Rabbit timing system~\cite{WhiteRabbit2011}, in order to \nima{provide} an absolute time-stamp for the camera triggers.  The White Rabbit system, developed initially at CERN, is an ``Open Hardware'' development for sub-nanosecond accuracy time-transfer, along with data transfer, by extending Synchronous ethernet with PTP (Precision Time Protocol)~\cite{PTP}.  The TiCkS board is connected to a commercial White Rabbit
\nc{Ethernet switch by a \sapoB{dedicated} fibre, which is used for data and timing signal transmission in both directions at different wavelengths.}
% through a single-mode optical fibre, through which data and timing signals are sent bidirectionally, using separate wavelength in each direction (which allows the transit time to be monitored continually).  
The White Rabbit Switch \nima{receives} absolute time from NTP (Network Time Protocol) combined with GPS time (Global Positioning System), and this time is transported to the White Rabbit nodes' internal clocks. 

\nima{The TiCkS internal clock is therefore continually kept at this distributed absolute time to within the CTA requirement of $1~\rm{ns}$ RMS relative to other CTA telescopes,  thanks to the White Rabbit protocol.}  When it receives camera trigger signals from the TIB, over two low-voltage differential signalling pairs -- for read-out or busy events\footnote{An event is flagged as busy when the TIB cannot send triggers to the FEBs because they are digitizing an event.}, respectively -- it increments the respective event counter and stores the absolute time-stamp of that camera trigger, along with the trigger-class and other additional information on the event provided by the TIB over an SPI link \nima{(Serial Peripheral Interface)~\footnote{\url{https://onlinedocs.microchip.com/pr/GUID-835917AF-E521-4046-AD59-DCB458EB8466-en-US-1/index.html?GUID-E4682943-46B9-4A20-A62C-33E8FD3343A3}}}.  The resulting trigger, counter, and camera information is sent in bunches as UDP \nima{(user datagram protocol)\footnote{\url{https://www.rfc-editor.org/rfc/rfc768.txt}}} packets  over the White Rabbit fibre, and which are then combined with the \sapoB{camera} pixel data based on the common event counters.  This is carried out in the \sapoB{camera event builder} process running on the \sapoB{camera server}. % (do we talk about the farm somewhere?). 

Also contained in the UCTS module is a \nima{Raspberry-Pi\footnote{\url{https://www.raspberrypi.org}}} based ``Remote Programming Board'' whose only function is to allow the firmware update of the TiCkS board over a JTAG link \nima{(Joint Test Action Group)~\cite{JTAG}}, by connecting through the standard Camera ethernet switches to the Raspberry-pi processor, without the need to physically access the UCTS module.

\begin{figure}
    \centering
    \includegraphics[width=\columnwidth]{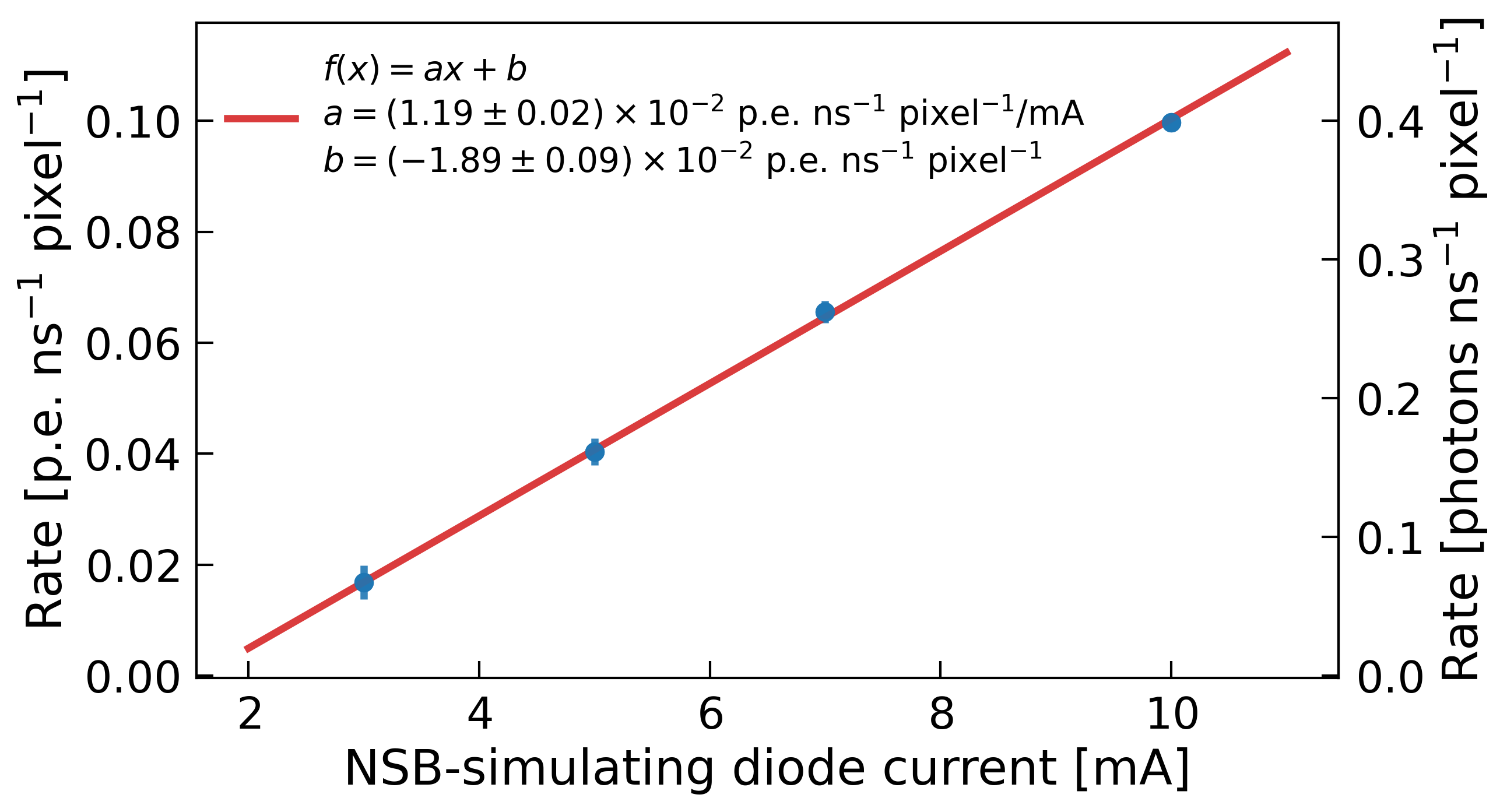}    
    \caption{Calibration of the NSB light source. The number of photoelectrons per ns and per pixel is plotted as a function of the \nima{NSB-simulating diode current amplitude}. The equivalent number of photons is shown on the right \nima{vertical axis. A dependence of NSB light emission on temperature is expected, with a measured variation of $\sim 10\%$ for a temperature gradient of $10^{\circ}$~C.} }
    \labfig{nsb}
\end{figure}

\section{Camera test setup}

The first NectarCAM unit has been integrated at the integration and test facility of CEA Paris-Saclay. 
\nc{The test facility housing the camera consists of a $\sim 12$~m long \nima{temperature controlled} dark room equipped with all camera services and calibration light sources, connected to a control room in which the full data acquisition (DAQ), storage and run control systems are located.}
The camera is equipped with the complete readout electronics {{consisting of}} 265 modules (1855 pixels), and with an entrance window, as can be seen in \reffig{nectarcam}. {The readout data are sent by \sapoA{UDP} to a dedicated \sapoB{camera server}~\cite{daq} through four {pairs} of 10\,Gb/s optical links. 
% A fifth optical fiber pair is used to control all camera devices from the control room, either using the {open platform communications unified architecture (OPC UA)} protocol or by UDP, from a graphical user interface running on a separate camera slow control server.
\sapoA{A fifth optical fiber pair is used to control all camera devices from the control room using a graphical user interface running on a separate camera slow control server.}

% During operation, the temperature of the darkroom is stabilized using ...... cooling system BLA BLA BLA --$>$ ask someone in the testbench!!

% During operation, the interior temperature of the camera is stabilised using water-air heat exchangers connected to an external, programmable thermocirculator. Humidity inside the camera is kept low using a constant inflow of dry air to avoid condensation, particuarly at low temperatures.
\nima{Three light sources are used to evaluate the camera at realistic intensities between roughly 1 and 12000~photons per pixel, corresponding to 0.25 and 3000~photoelectrons (p.e.) after conversion in the PMTs (using 25\% collection efficiency). \nimabis{The charge versus number of photoelectrons measurement is linear to a very good approximation \cite{linearity}}. The three light sources are a flat field calibration pulsed light source (FFCLS), a continuous night sky background (NSB) emulating light, and a pulsed laser source. The FFCLS and the NSB sources are also controlled from the graphical interface in the control room. Once the camera is on-site, the FFCLS and the NSB sources are going to be used for calibration. However, a laser source is also used in the dark room to perform precise timing measurements.}

%% FFCLS
The FFCLS consists of 13 light emitting diodes (LEDs), emitting pulsed light at 390~nm wavelength \nima{(FWHM of 3.3~ns)} to reproduce {the maximum of the received UV} Cherenkov spectrum. A holographic filter is mounted in front of the LEDs in order to obtain a diffuse and uniform light emission on the camera. The intensity of the FFCLS can also be adjusted using transmission filters. 
An optical fiber connects the FFCLS directly to the TIB of the camera to give the possibility to trigger externally.
Moreover, the flat-field calibration source can be triggered \nima{in internal mode} periodically  or by an external random generator. \nima{The random generator is} made with a stable light source and \nima{a H12386 Hamamatsu photon counting head device\footnote{\url{https://www.hamamatsu.com/content/dam/hamamatsu-photonics/sites/documents/99_SALES_LIBRARY/etd/H12386_TPMO1073E.pdf}}, which consists of a photomultiplier tube, a high-speed photon counting circuit, and a high-voltage power supply circuit.} \nima{The photon counting output provides a random arrival time which is used for the generation of a random trigger.} \nima{The uniformity of the FFCLS light emission across the camera has been measured to be better than 2\%.}
% \nc{\st{The flatness of the FFCLS was evaluated for the full camera scale with the help of a movable x-y table, {resulting} so measured} to be better than 2\% \st{level}.}

%% NSB
The NSB light source is used in order to reproduce the \nima{median photoelectron rate} of the typical night sky {spectrum}. The light is created by a 519~nm LED in combination with a tunable diaphragm, resulting in a light output with {spatial uniformity at the camera level} better than 5\% \sapoB{RMS}. The NSB source has been calibrated by illuminating the camera with a low energy flux and counting the number of photons in the full 60~ns waveform length for each camera pixel. The result of the calibration is shown in \reffig{nsb}, where the number of photoelectrons measured per nanosecond and pixel is plotted as a function of the NSB current.

% peak wavelength between 500 nm and 600 nm (green);
% • the spatial uniformity of the ”light intensity” better than 5%. This quan-
% tity will be interpreted as an irradiance;
% • ”intensity level” between 0.24 ph/(ns.sr.cm2) and 1.2 ph/(ns.sr.cm2). This quantity will be interpreted as a radiance;
% • atunableintensityrangefrom0.01photo-electron(p.e.)to1photo-electron per ns and per illuminated pixel.

\begin{figure}[t!]
    \centering
    \includegraphics[width=\columnwidth]{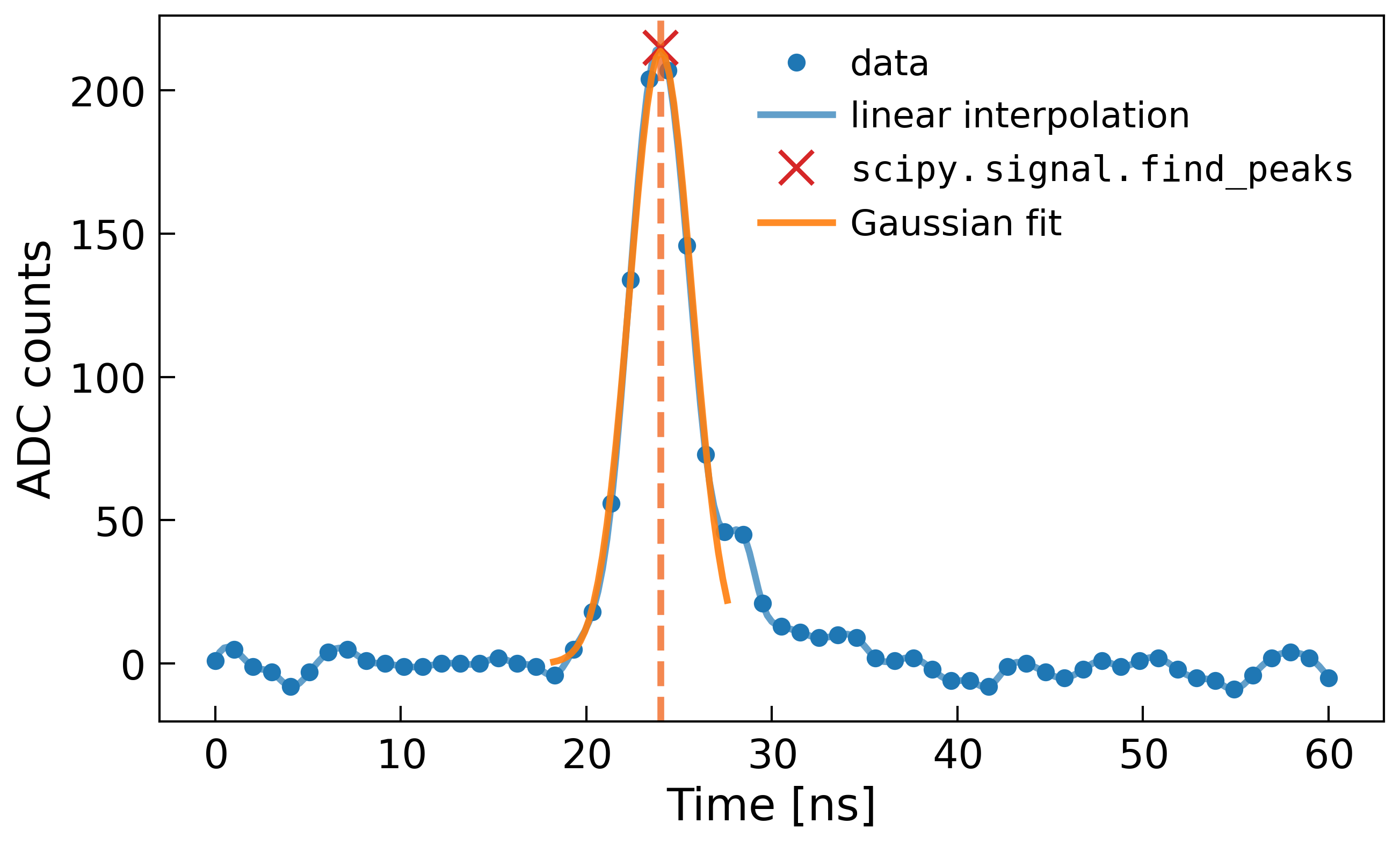}
    \caption{Waveform for one event after pedestal subtraction. The data points are interpolated using the \nima{\texttt{scipy.interpolate.InterpolatedUnivariateSpline}} function (blue line). The two methods used to identify the TOM are shown. The red cross shows the maximum of the peak found with the \texttt{signal.find\_\_peaks} \texttt{scipy} function\nc{~\cite{scipy}}. The orange curve represents the Gaussian fit. The vertical orange and red dashed lines show the TOM positions found with the first and second method, respectively. \nima{The red line is not visible since it coincides with the orange one.}}
    \labfig{tom_waveform}
\end{figure}
% laser: https://www.picoquant.com/products/category/picosecond-pulsed-sources/ldh-i-series-smart-laser-diode-heads-for-taiko-pdl-m1
The laser source is a pulsed diode \nimabis{(Picoquant LDH-IB-375-B\footnote{\url{https://www.picoquant.com/images/uploads/downloads/18451-ldh-i_series.pdf}}, $50$~ps width)} creating a uniform illumination of the full camera at a wavelength of 373~nm. \nimabis{The light from the laser is attenuated and sent to the camera through a diaphragm. The total attenuation was measured to be 38.}
The laser is also connected to a \sapoB{UCTS board} that records the timestamp of the \nima{photon laser pulse} through \sapoB{a TiCkS board}, {allowing to perform precise measurements of the camera timing \nima{performance}.} {According to the manufacturer, the jitter of the laser is $<20$~ps (FWHM), therefore it is going to be considered \nc{negligibly} small in the rest of the paper.}
% The laser is also connected to a \sapoB{UCTS board} that records the timestamp of the photon through a \nc{customised} White Rabbit based board {(TiCkS)}, {allowing to perform precise measurements of the camera timing performances.} {According to the manufacturer, the jitter of the laser is $<20$~ps (FWHM), therefore it is going to be considered \nc{negligibly} small in the rest of the paper.}

% laser model is  IB-375-B
All light sources are at a distance of 12~m from the center of \sapoB{NectarCAM}.

\section{Time of Maximum}
\labsec{tom}
The temporal information of a PMT signal is sampled every nanosecond for {a total window} of 60~ns. The reconstructed arrival time of the pulse signal is given by the temporal position of the pulse maximum in the sampled window. This variable is called time of maximum (TOM).

The TOM is estimated from the waveform after subtracting the pedestal. Two methods have been used (see \reffig{tom_waveform}). The first consists in identifying the position of the largest peak of the waveform using the function \texttt{signal.find\_peaks} from the \texttt{scipy} python package~\cite{scipy}. The result was then cross-checked by performing a Gaussian fit of the largest peak of the waveform using as \nima{initialization value} the position of the peak obtained from the first method.

\section{{Single pixel timing precision}}

% \section{{Single pixel timing resolution}}
% Pixel systematic timing uncertainty}
According to CTA requirements, NectarCAM needs to have a single pixel timing precision better than 1~ns for a light illumination larger than 20~photons (i.e., 5~p.e.). To estimate the single pixel timing precision, we illuminate the camera with a uniform light created by the laser source at a frequency of 1~kHz with \nimabis{pulsed energies between 8~pJ and 20~pJ}, and we measure the TOM of each pulse for each pixel. \nimabis{Laser energies per pulse of 8~pJ and 20~pJ correspond to expected numbers of photons of $\sim 1$~p.e. and $\sim 200$~p.e. per pixel, respectively.} \nimabis{The 2.5-fold difference in laser power displayed on the laser driver corresponds to a 200-fold difference in photoelectrons measured in the camera due to the non-linearity of the laser output.} The \nima{TOM of individual events} has been calculated using both methods described in \refsec{tom} \nc{and are \sapoA{in agreement within 50~ps}}. \nima{As explained later, the dispersion of the TOM of events for a given pixel gets a contribution from the random starting time of the NECTAr chip readout.} The systematic timing uncertainty of each pixel can be estimated by the root mean square (RMS) of the obtained \nc{TOM} distributions.
\reffig{pixel_res} shows the weighted mean of the RMS over all the pixels as a function of the illumination charge. The results from both methods are shown. The weight is given by the inverse of the \nima{square of the} standard deviation of the \nc{TOM} distribution of each pixel.
{For an incoming light of intensity above $\sim10$~photons,} both methods show that the time precision is \nima{better} than 1~ns, fulfilling the CTA requirement \nc{according to which the RMS uncertainty on the mean relative reconstructed arrival time in every pixel must not exceed 1~ns for amplitudes in the range 20 to 2000 photons (see violet dashed line and area).} However, already at a light illumination of 800~photons, the pixel time precision {{limit}} is reached, as shown by the gray dashed line. \sapoA{This limit is due to the \nima{quantization} of the L1A trigger acceptance signal by the NECTAr chip. \nima{This is due to the asynchronous arrival of the LA1 trigger with respect to the NECTAr chip clock.} Since the NECTAr chip samples at 1~GHz, this corresponds to an RMS of $1/\sqrt{12}~\mathrm{ns} =290$~ps.} }

% This limit \nc{corresponds to} the {quantification noise due to the NECTAr 1~GHz sampling and corresponds to an RMS of $1/\sqrt{12}~\mathrm{ns} =290$~ps.}

%Actually this limit is not related to the quantization step for the digitized data (the interpolation allows to reconstruct the peak position with a precision better than a step). It is indeed related to the timing precision of the capture of the digital L1a signal by NECTAR. This is the critical operation as it defines the index of the first cell to read and thus the t0. This is this capture which is affected by the quantization step.

\begin{figure}[t!]
    \centering
    \includegraphics[width=\columnwidth]{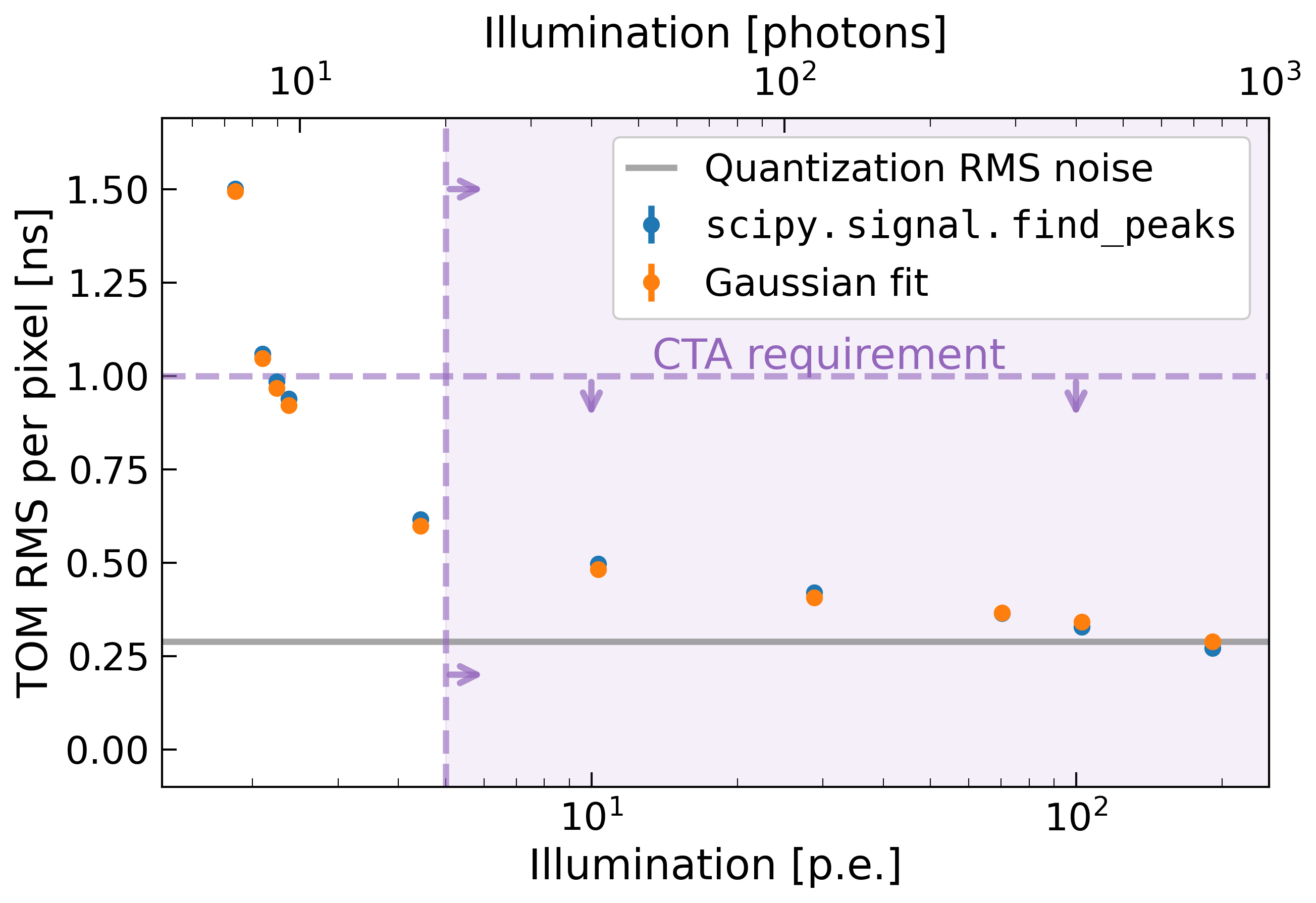}
    \caption{Timing precision per pixel (in ns) as a function of the charge of the illumination signal (in photoelectrons and photons on the bottom and top of the \nima{horizontal} axis, respectively). The timing resolution is given by the mean of the RMS distribution over all the 1855 pixels. Both methods are shown (in blue and orange). The gray \nima{solid} line shows the \nima{quantization} {(RMS)} noise given by $\frac{1}{\sqrt{12}}$~ns. \sapoA{The dashed violet lines and arrows show} the 1~ns requirement limit to be valid between 20[5] and 2000[500]~photons~[p.e.] (violet area).}
    \labfig{pixel_res}
\end{figure}
\begin{figure*}[t!]
    \centering
    \includegraphics[width=\textwidth]{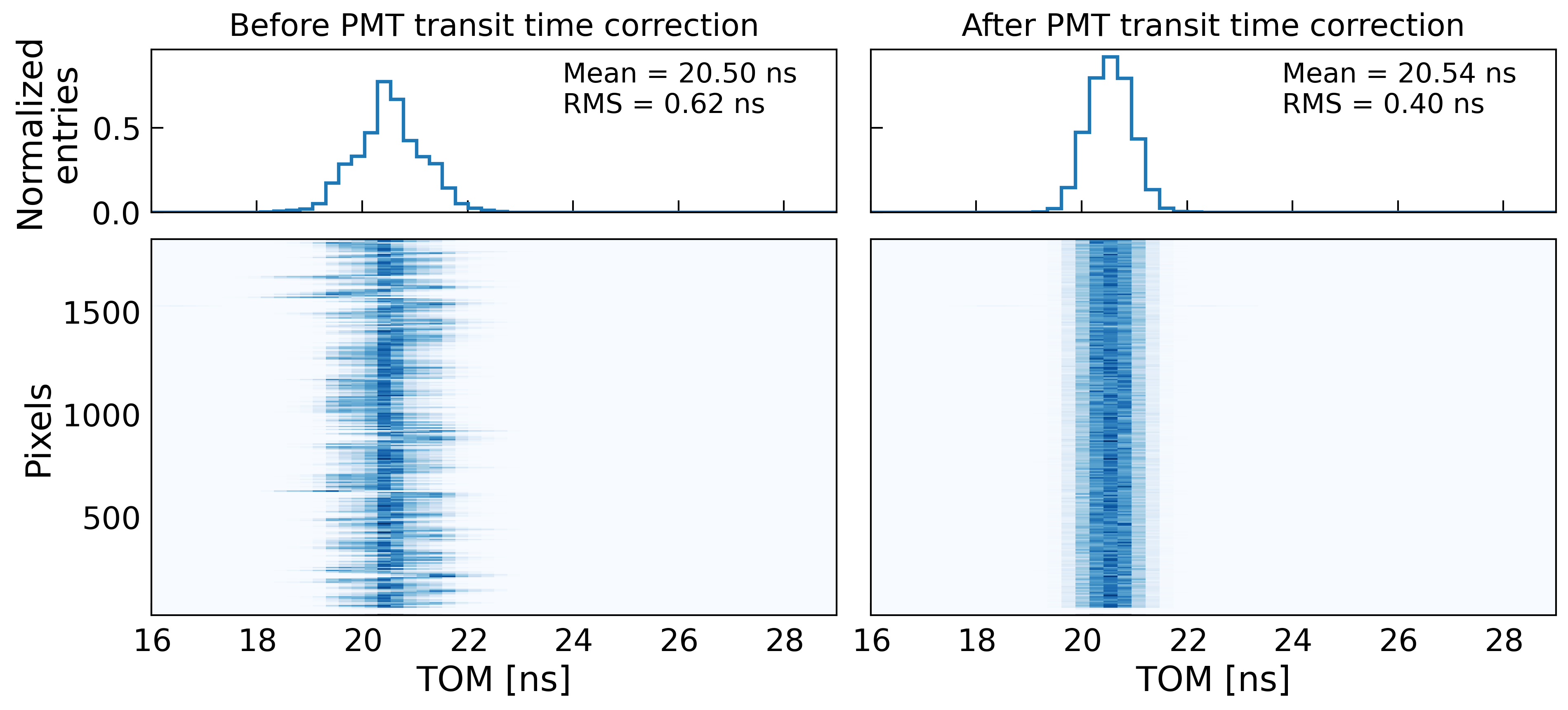}

    \caption{\nima{Mean TOM distribution for all pixels before (left plot) and after (right plot) the PMT transit time correction for a light illumination of $\sim 70$~p.e. and for a  uniform high voltage of 1000~V for all the pixels.}}
    \labfig{pmt_tt}
\end{figure*}

\section{PMT transit time}
\labsec{pmt_tt}
Prior to the estimation of the camera {global} timing precision, it is necessary to perform a {relative} calibration of the \nc{TOM} over all the 1855 pixels. Two systematic effects \nima{combine} to generate offsets. 
The first one is caused by the different arrival times of a L1A trigger signal from the TIB to the modules due to a different jitter in the FPGA of each DTB~\cite{ICRC2019}.
The L1A signal is sent by the TIB to the front end modules to stop the GHz sampling of the NECTAr chips and start acquiring the waveform data. 
{The time dispersion due to L1A delay \nima{between modules} can be corrected by tuning a delay in each digital \sapoB{backplane}. The method is similar to the more complex PMT transit time calibration, which will be described in this paragraph, except that the former can be corrected during data taking {while the latter can only be corrected at the analysis stage}.} %between modules can be taken into account with calibration. However, it is not an issue for the measurement discussed in this paper and will be addressed in a future work.}

{The second effect is the PMT transit time, which is the transfer time of the electron {avalanche} in the PMT~\cite{leo} and depends on the high voltage applied to the dynodes. 
However, the high voltages also determine the amplitude of the output pulses of the PMTs. Therefore, since they are selected for each pixel in order to have the same nominal gain of 40000 over all the camera detection plane, the PMTs will introduce different delays, typically degrading the trigger \nima{performance}. The drop in telescope \sapoB{performance} in terms of effective area due to the PMT transit time becomes especially noticeable for gamma-ray energies smaller than 100~GeV  \cite{pmt_tt_simulations}.
However, this time offset between pixels can be corrected at the analysis level in order to improve the time resolution of the camera.}
% Each pixel works at a different high voltage in order to have the same nominal gain of 40000 over all the camera detection plane. This causes a time offset between pixels that can be corrected at the analysis level to improve the time resolution of the camera.
The effect of the transit time spread between pixels is illustrated by the left panel of \reffig{pmt_tt}, showing that the mean \nc{TOM} distributions for all pixels do not overlap but are shifted in time. % when illuminating the camera with the 13 LEDs of the FFCLS at {{?? V}}.
\begin{figure}[t]
    \centering   
    \includegraphics[width=\columnwidth]{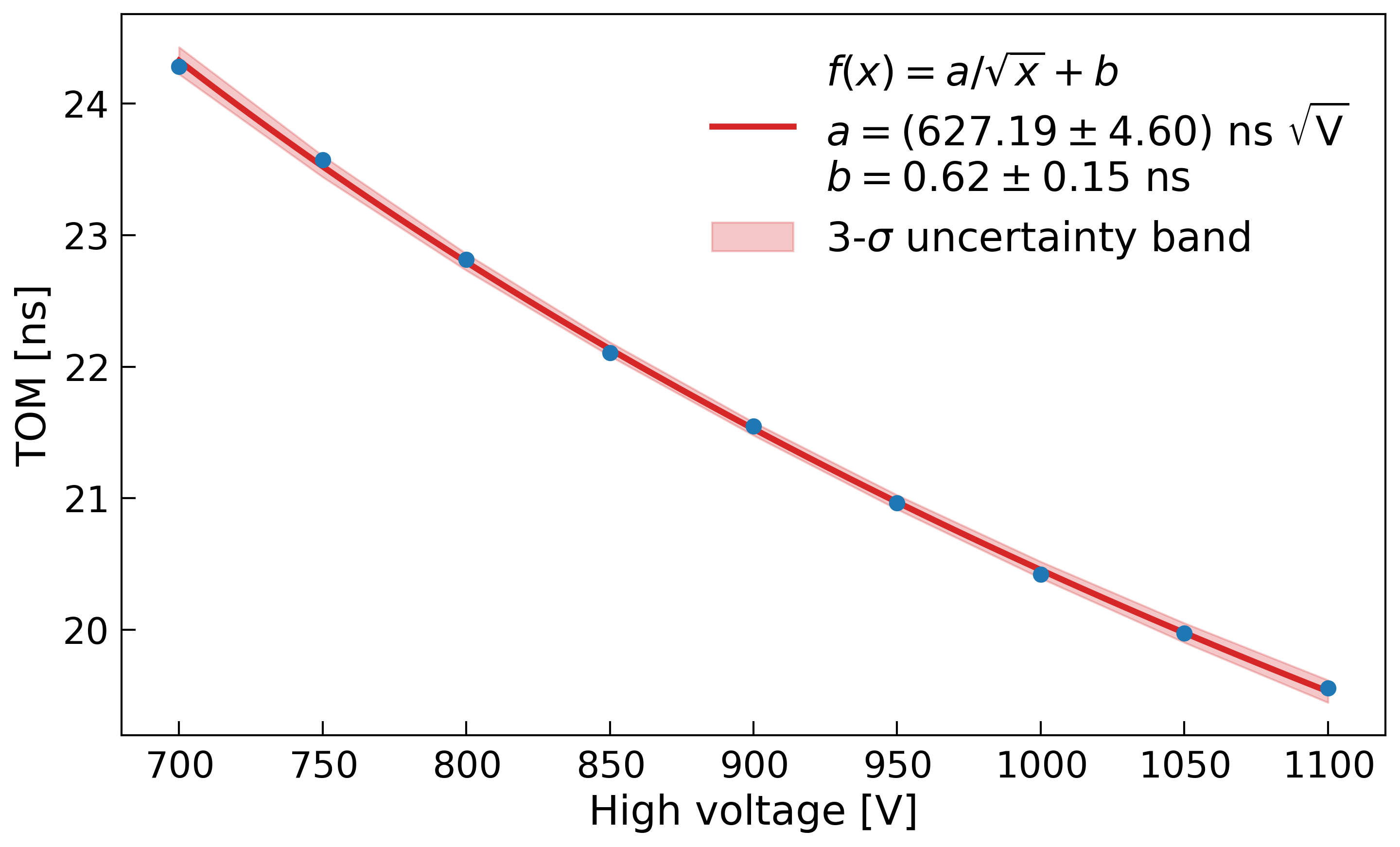}
    \caption{PMT transit time versus voltage for one pixel. The \nima{data points shows the mean TOM of $\sim$1000 events each} in the camera as a function of the high voltage applied to all PMTs. The least-squares fit with the $3-\sigma$ uncertainty band is shown in red.}
    \labfig{pmt_tt_fit}
\end{figure}

In order to correct for the PMT transit time effect,  all pixels in the camera have been set to the {same voltage between 700~V and 1100~V} and illuminated with different intensities of the 13 LEDs of the FFCLS. For each light intensity, we have calculated the average \nc{TOM} for each pixel using the \texttt{scipy.signal.find\_peaks} method described in \refsec{tom}. {A least-squares method fit has been performed for all pixels, using the function $f(x)=a/\sqrt{x} +b$, where $x$ are the PMT voltages. This function parametrizes the time required by an electron to be accelerated by the electric field of the PMT.}
% A linear fit of these values as a function of the FFCLS voltages has been then performed for all pixels. 
\reffig{pmt_tt_fit} shows the fit for a single pixel. \nima{The data points show} the mean of the \nc{TOM} over $\sim$~1000 events. The results of the fit of each pixel are then used to correct for the PMT transit time effect. \nima{After calculating the fit value for each pixel at 1000~V, the TOM is shifted to the overall average of all pixel values at the same voltage.} The result of the correction is shown in the right panel of \reffig{pmt_tt}, where the mean \nc{TOM} distributions of {all pixels} are now synchronized. {The \nc{quality} of the PMT transit time correction is shown by the RMS of the distribution the mean TOM for all pixels which is reduced from \nima{0.62~ns} before correction to \nima{0.40~ns} after correction.}
% new wrt to V0
For each pixel, the fit parameters and the PMT transit time correction with respect to the average value at 1000~V are saved in a {database}, which will be used in the future for the calibration of the camera. In this way, if the nominal voltage of each pixel {changes}, we will be able to recalibrate the pixel timing {to correct for} the PMT transit time effect. {In fact, the correction of the PMT transit time effect will take place offline, by shifting the measured TOM for each pixel to the corrected one using the aforementioned fit parameters.}
\nima{Due to PMT aging, the high voltage of each PMT will be adjusted to the nominal gain. Therefore, the PMT transit time is expected to change during operation time.}

\begin{figure}[ht]
    \centering
    \includegraphics[width=\columnwidth]{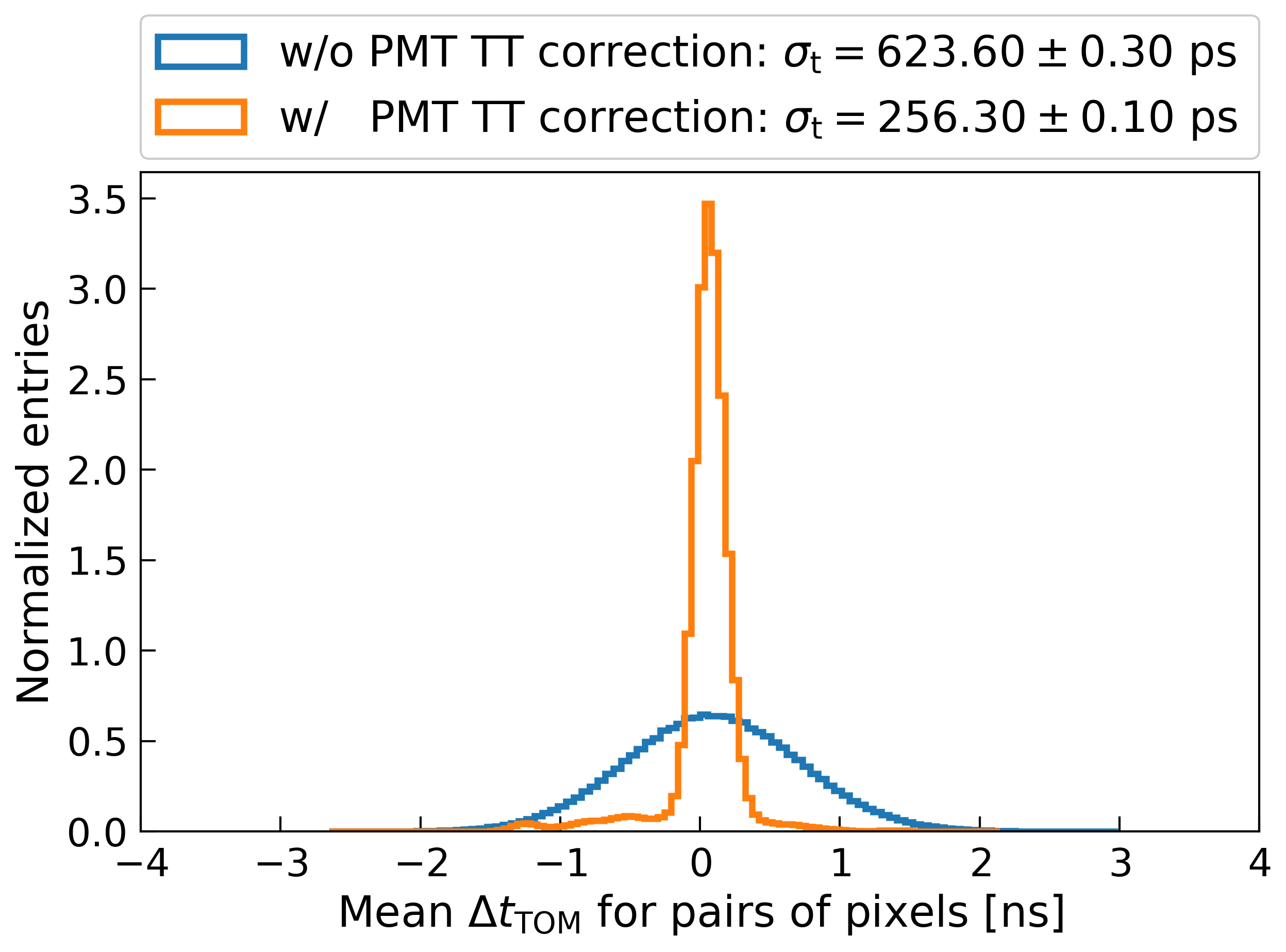}
    \caption{\sapoA{Normalized distribution of the mean difference between the \nc{TOM} value for each couple of pixels over all the events. The distributions with and without PMT transit time correction are shown in orange and blue, respectively. The standard deviation $\sigma_t$ of the two distributions is shown in the legend \nima{and the error is only statistical.}.}}
    \labfig{delta_t_pixels}
\end{figure}

\section{{Global camera  timing precision}}
After correcting for the PMT transit time, the timing resolution of the camera can be evaluated. This measurement is performed by illuminating the camera with a uniform light at $\sim 20$~p.e. generated by \sapoA{the} laser source. 
{Since the measurement is performed \sapoA{in internal trigger mode\footnote{The internal trigger mode is the standard trigger for cosmic data taking.}}, the $\Delta t_{\mathrm{TOM}}$ calculation already takes into account the difference in distances travelled by the light before reaching \sapoA{any two pixels pair in the whole camera}, no matter the position of the pixels in the camera. \nima{This is because the laser light source which has travel time differences between pixels is also used to perform the L1A calibration.} Therefore, no geometrical correction is needed.} The \nc{TOM} difference $\Delta t_{\mathrm{TOM}}$ between {all pairs} of pixels is then calculated with and without PMT transit time correction. The difference is calculated for each pair of pixels and for all events in the dataset. The final mean distribution over all events for each pair of pixels is plotted in \reffig{delta_t_pixels}, with and without PMT transit time correction. 
\nima{After calibration, the mean of the TOM difference distribution for any two simultaneously illuminated pixels is reduced from $623.6\pm0.3$~ps to $256.3\pm0.1$~ps}. \nima{Moreover, the RMS distribution of the $\Delta t_{\mathrm{TOM}}$ for pair of pixels can as be computed, as shown in \reffig{delta_t_rms}. This distribution is below the 2~ns CTA requirement on the global camera timing precision.}

\begin{figure}
    \centering
    \includegraphics[width=\columnwidth]{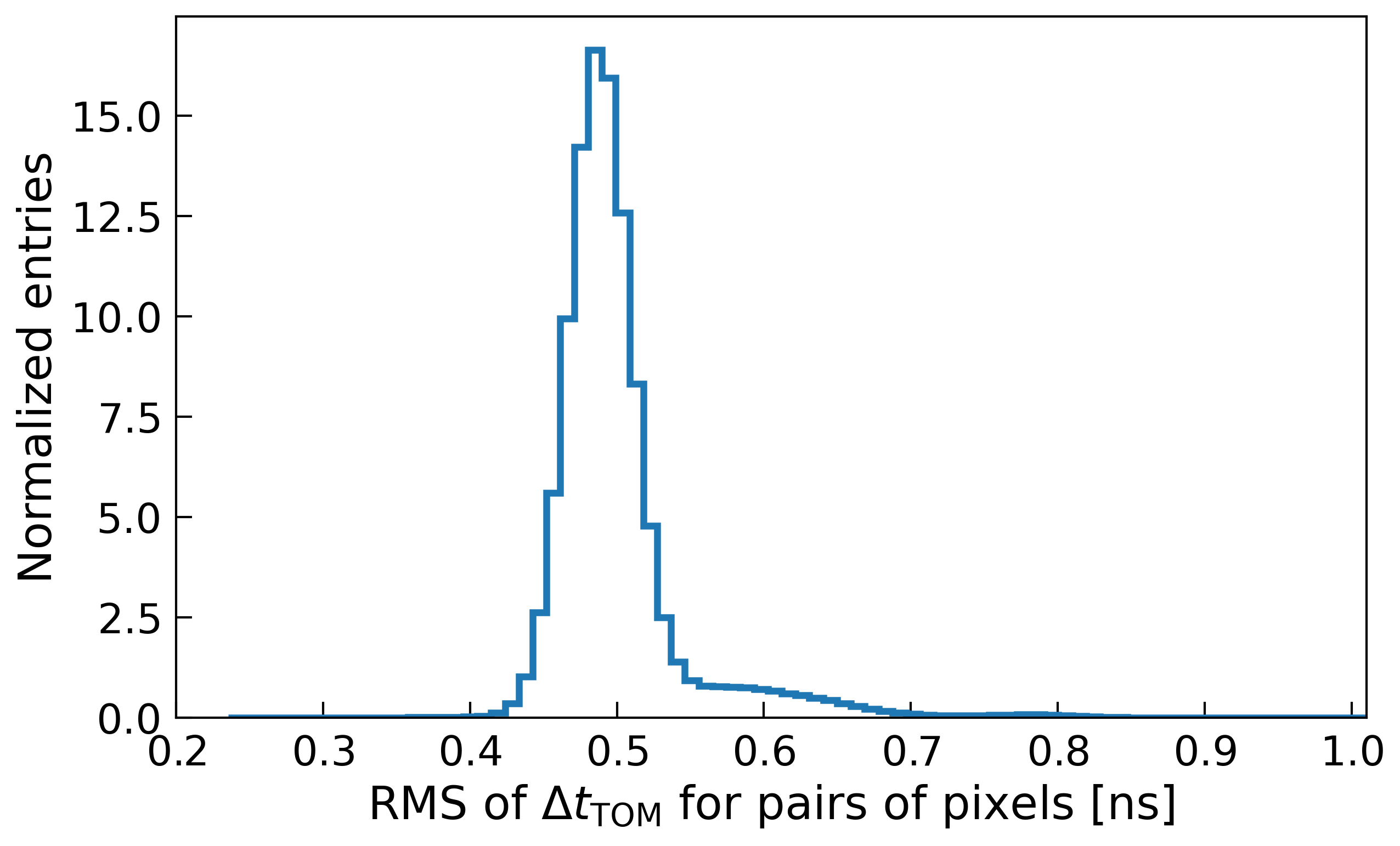}
    \caption{RMS distribution of the \nc{TOM} difference between all \nima{pairs} of pixels after applying the PMT transit time correction. }
    \labfig{delta_t_rms}
\end{figure}

\section{Camera trigger timing accuracy}
{The camera events are timestamped in the TiCkS board of the UCTS module. The timestamp will then \nima{be} used by the CTA central trigger process to correlate the triggers of the different telescopes of the CTA array. The trigger time is required to be measured with a 2~ns accuracy.} %To quantify the RMS of the timestamp distribution, we have used two different methods.}

{To quantify the RMS of the timestamp distribution, we have illuminated the full camera using the laser source at 1~kHz with different intensities ranging from $\sim 20$~p.e. up to \nima{$\sim190$~p.e.} For each measurement, the start time of the laser flashes are recorded with an external TiCkS board. In this way, it is possible to measure directly the time difference between the light arrival in the detection plane and the output of the timestamp to the CTA central trigger ($\Delta t_{\mathrm{TiCkS}} = t_{\mathrm{UCTS}} - t_{\mathrm{laser}}$). This method has been compared with the distribution of the time difference of \nc{two} consecutive events, which gives an upper limit on the accuracy of the timestamps for a {perfectly} periodic input signal. Given the 1kHz frequency, the time difference between two consecutive camera trigger events $\Delta t_{\mathrm{UCTS}} = t_{\mathrm{UCTS},i} - t_{\mathrm{UCTS}, i-1}$ is expected to be $\sim 1 \times 10^6$~ns, {as shown in \reffig{delta_ucts}.} The pixel charge for every trigger is analyzed in order to \nima{filter out} the cosmic rays events, producing a $\Delta t_{\mathrm{UCTS}}$ histogram from which the RMS uncertainty is obtained. \reffig{trigger_res} shows the RMS values for every configuration used.} 
The resulting trigger timing accuracy is better than 0.5 ns with both methods, and thus below the CTA requirements of 2~ns.

% The first experimental setup consists in illuminating the camera with the FFCLS using 13 LEDs at $\sim7$~p.e. and a frequency of 1~kHz in internal trigger mode with a 5~p.e. threshold. {For this setup, the distribution of the time difference of 2 consecutive events for a periodic input signal gives an upper limit on the accuracy of the timestamps.}
% For the second setup, the laser source is also set at 1~kHz with different intensities ranging from $\sim 2$~p.e. up to $\sim50$~p.e. with a TiCkS board. {With this setup, it is possible to measure directly the time difference between the light arrival in the detection plane and the output of the timestamp to the CTA central trigger.}

% Given the 1kHz frequency, the time difference between two consecutive camera trigger events $\Delta t_{\mathrm{UCTS}} = t_{\mathrm{UCTS},i} - t_{\mathrm{UCTS}, i-1}$ is expected to be of $1 \times 10^6$~ns. The pixel charge for every trigger is analyzed in order to filter the cosmic rays events, producing a $\Delta t_{\mathrm{UCTS}}$ histogram from which the RMS uncertainty is obtained. \reffig{trigger_res} shows the RMS values for every configuration used.

\begin{figure}[t]
    \centering
    \includegraphics[width=\columnwidth]{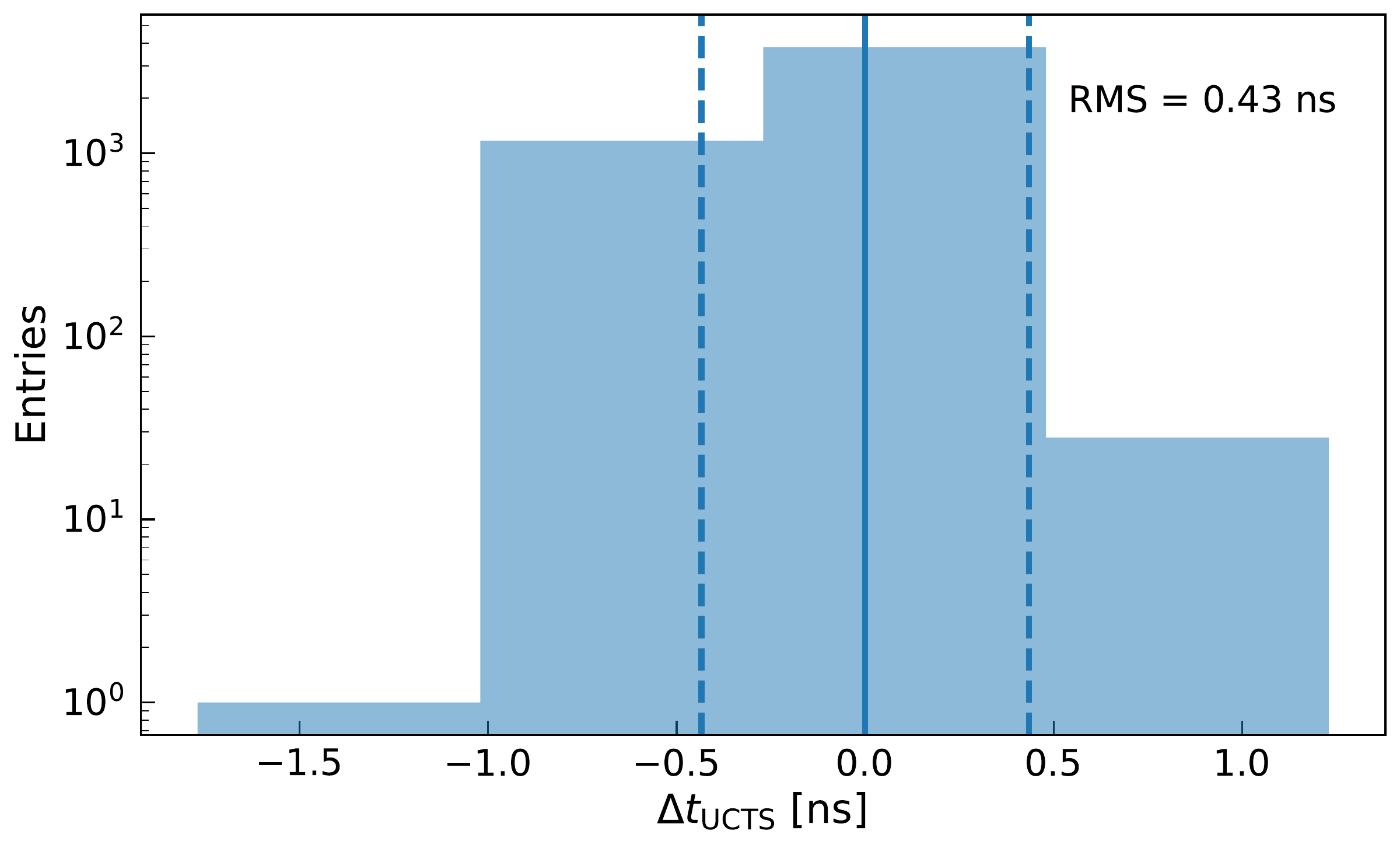}
    \caption{Distribution of the UCTS timestamp differences between two consecutive events for an illumination of 191[764]~p.e.[photons] \nima{after subtracting the mean value of 1~ms.} The solid and dashed lines show the mean and RMS, respectively. }
    \labfig{delta_ucts}
\end{figure}

\section{Conclusions}
The first unit of \sapoB{NectarCAM} has been integrated at the integration and test facility of CEA Paris-Saclay. The first timing performance verifications have shown that the camera fulfills the CTA requirements. \nima{For the test, two pulsed light sources have been used prior to calibration: a flat-field calibration light source and a laser.}
\nima{The single pixel timing precision has been measured to be less than 1~ns for an incoming light of intensity between $\sim~10$ and $\sim~1000$ photons, making it suitable for CTA usage.} %demand
% The global camera timing precision has also been studied. It consists in analyzing the time difference in the time pulse reconstruction between pairs of pixels. 
\sapoB{The global camera timing precision has also been studied by analyzing the difference in pulse-time reconstruction between pairs of pixels.} \nima{After correcting for the PMT transit time of each pixel, the time precision of the camera is reduced to 0.5~ns (\reffig{delta_t_rms}), in line with the CTA demand.} %from \nima{$0.62$~ns to $0.26$~ns}. 
Furthermore, the trigger \nima{performance has} been tested.  The timing accuracy that comes from the trigger timestamp of the camera relative to the time of arrival of light at the detection plane has been measured to be below 0.5~ns \nima{for the illumination range between  $\sim20$~p.e. and $\sim190$~p.e., using two different methods. Therefore, it complies with the CTA requirements.}
% We have measured this uncertainty by illuminating the camera with a laser and a flat-field source, both in combination with an NSB source, and by using two methods. In all cases, the time uncertainty on the UCTS timestamp of the camera is of the order of tenths of ns, and thus compliant with the CTA requirements. 
\begin{figure}[t]
    \centering
    \includegraphics[width=\columnwidth]{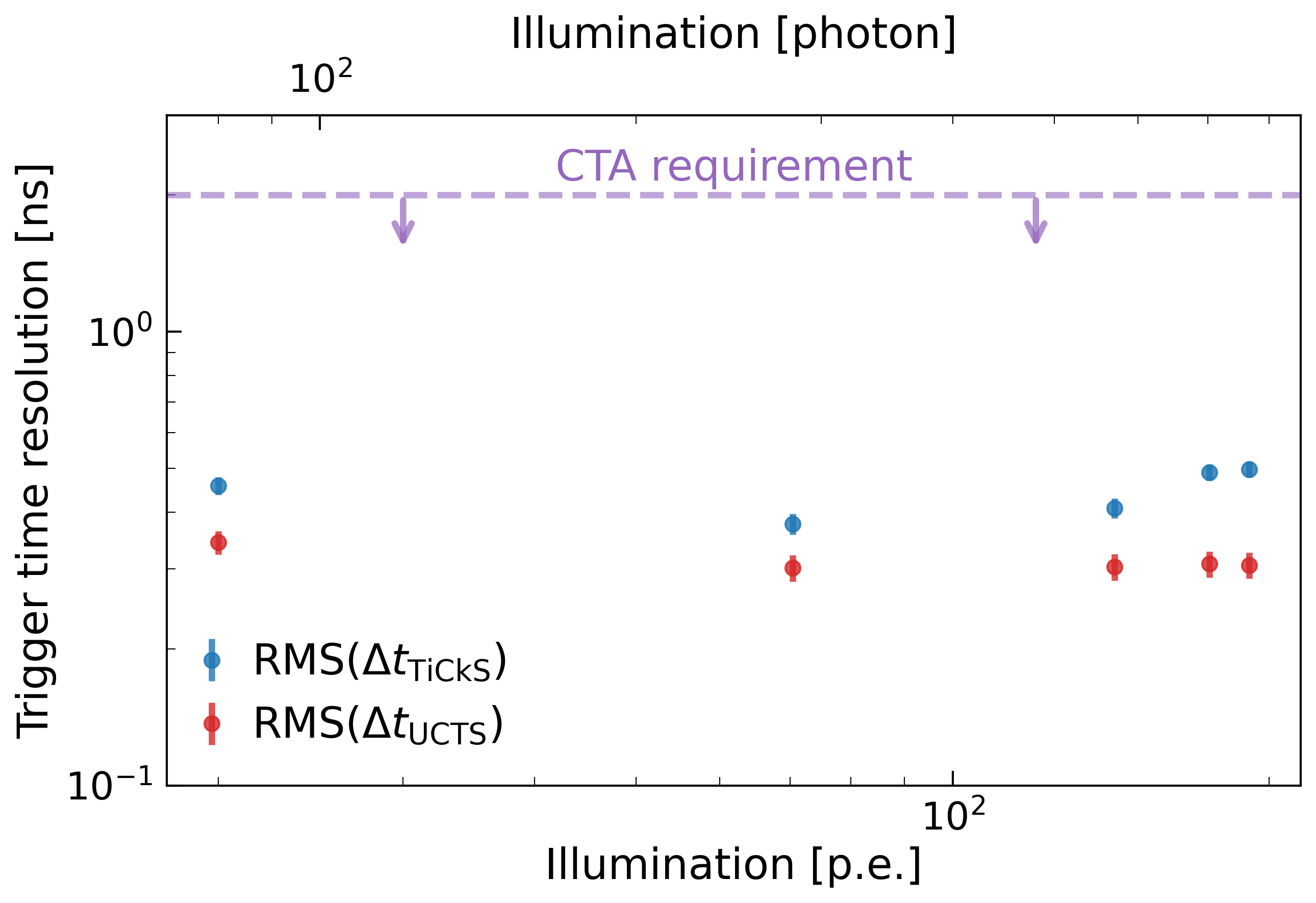}
    \caption{Camera trigger precision in ns as a function of the charge of the illumination signal (in photons and photoelectrons on the bottom and top of the \nima{horizontal} axis, respectively) using two different methods. {The CTA requirement of 2~ns for the camera timing resolution is also shown by the violet dashed line.} }
    \labfig{trigger_res}
\end{figure}

\section{Acknowledgements}
We thank the anonymous referees for carefully reading the manuscript and providing helpful suggestions that improved the clarity of the text. This work was conducted in the context of the CTA Consortium. We gratefully acknowledge financial support from the following agencies and organizations: Ministry of Higher Education and Research, CNRS-INSU and CNRS-IN2P3, CEA-Irfu, ANR, Regional Council Ile de France, Labex ENIGMASS, OCEVU, OSUG2020 and P2IO, France; DESY, Helmholtz Association, Germany; Spanish Research State Agency (AEI) through the grant PID2019-104114RB-C32.
\bibliographystyle{elsarticle-num}
\bibliography{mybibfile}

\begin{thebibliography}{10}
\expandafter\ifx\csname url\endcsname\relax
  \def\url#1{\texttt{#1}}\fi
\expandafter\ifx\csname urlprefix\endcsname\relax\def\urlprefix{URL }\fi
\expandafter\ifx\csname href\endcsname\relax
  \def\href#1#2{#2} \def\path#1{#1}\fi

\bibitem{hawc}
A.~U. Abeysekara, et~al., {Observation of the Crab Nebula with the HAWC
  Gamma-Ray Observatory}, Astrophys. J. 843~(1) (2017) 39.
\newblock \href {http://arxiv.org/abs/1701.01778} {\path{arXiv:1701.01778}},
  \href {https://doi.org/10.3847/1538-4357/aa7555}
  {\path{doi:10.3847/1538-4357/aa7555}}.

\bibitem{lhaaso}
F.~Aharonian, et~al., {The observation of the Crab Nebula with LHAASO-KM2A for
  the performance study}, Chin. Phys. C 45~(2) (2021) 025002.
\newblock \href {http://arxiv.org/abs/2010.06205} {\path{arXiv:2010.06205}},
  \href {https://doi.org/10.1088/1674-1137/abd01b}
  {\path{doi:10.1088/1674-1137/abd01b}}.

\bibitem{CTAconcept}
B.~Acharya, et~al., {Introducing the CTA concept}, Astroparticle Physics 43
  (2013) 3--18, {Seeing the High-Energy Universe with the Cherenkov Telescope
  Array - The Science Explored with the CTA}.
\newblock \href
  {https://doi.org/https://doi.org/10.1016/j.astropartphys.2013.01.007}
  {\path{doi:https://doi.org/10.1016/j.astropartphys.2013.01.007}}.

\bibitem{cta_mc_design}
K.~Bernl\"{o}hr, et~al., {Monte Carlo design studies for the Cherenkov
  Telescope Array}, Astroparticle Physics 43 (2013) 171--188, {Seeing the
  High-Energy Universe with the Cherenkov Telescope Array - The Science
  Explored with the CTA}.
\newblock \href
  {https://doi.org/https://doi.org/10.1016/j.astropartphys.2012.10.002}
  {\path{doi:https://doi.org/10.1016/j.astropartphys.2012.10.002}}.

\bibitem{nectarcam}
J.-F. Glicenstein, {NectarCAM : a camera for the medium size telescopes of the
  Cherenkov Telescope Array}, PoS ICRC2015 (2016) 937.
\newblock \href {https://doi.org/10.22323/1.236.0937}
  {\path{doi:10.22323/1.236.0937}}.

\bibitem{2021NIMPA100765413T}
A.~{Tsiahina}, et~al., {Measurement of performance of the NectarCAM
  photodetectors}, Nuclear Instruments and Methods in Physics Research A 1007
  (2021) 165413.
\newblock \href {https://doi.org/10.1016/j.nima.2021.165413}
  {\path{doi:10.1016/j.nima.2021.165413}}.

\bibitem{lightconcentrators}
F.~{H{\'e}nault}, et~al., {Design of light concentrators for Cherenkov
  telescope observatories}, in: R.~{Winston}, J.~{Gordon} (Eds.), Nonimaging
  Optics: Efficient Design for Illumination and Solar Concentration X, Vol.
  8834 of Society of Photo-Optical Instrumentation Engineers (SPIE) Conference
  Series, 2013, p. 883405.
\newblock \href {https://doi.org/10.1117/12.2024049}
  {\path{doi:10.1117/12.2024049}}.

\bibitem{acta}
D.~{Gascon}, et~al., {Wideband pulse amplifier with 8 GHz GBW product in a 0.35
  {\ensuremath{\mu}}m CMOS technology for the integrated camera of the
  Cherenkov Telescope Array}, Journal of Instrumentation 5~(12) (2010) C12034.
\newblock \href {https://doi.org/10.1088/1748-0221/5/12/C12034}
  {\path{doi:10.1088/1748-0221/5/12/C12034}}.

\bibitem{nectar0}
E.~Delagnes, et~al., {NECTAr0, a new high speed digitizer ASIC for the
  Cherenkov Telescope Array}, in: 2011 IEEE Nuclear Science Symposium
  Conference Record, 2011, pp. 1457--1462.
\newblock \href {https://doi.org/10.1109/NSSMIC.2011.6154348}
  {\path{doi:10.1109/NSSMIC.2011.6154348}}.

\bibitem{trigger}
U.~Schwanke, et~al., {A versatile digital camera trigger for telescopes in the
  Cherenkov Telescope Array}, Nuclear Instruments and Methods in Physics
  Research Section A: Accelerators, Spectrometers, Detectors and Associated
  Equipment 782 (2015) 92--103.
\newblock \href {https://doi.org/https://doi.org/10.1016/j.nima.2015.01.096}
  {\path{doi:https://doi.org/10.1016/j.nima.2015.01.096}}.

\bibitem{L0}
D.~{Gascon}, et~al., {Reconfigurable ASIC for a low level trigger system in
  Cherenkov Telescope Cameras}, Journal of Instrumentation 11~(11) (2016)
  P11017.
\newblock \href {https://doi.org/10.1088/1748-0221/11/11/P11017}
  {\path{doi:10.1088/1748-0221/11/11/P11017}}.

\bibitem{3nn_mc}
T.~{Armstrong}, et~al., {Monte Carlo Simulations and Validation of NectarCAM, a
  Medium Sized Telescope Camera for CTA}, in: 37th International Cosmic Ray
  Conference, 2022, p. 747.
\newblock \href {https://doi.org/10.22323/1.395.0747}
  {\path{doi:10.22323/1.395.0747}}.

\bibitem{KAPUSTINSKY1985612}
J.~Kapustinsky, R.~DeVries, N.~DiGiacomo, W.~Sondheim, J.~Sunier, H.~Coombes,
  \href{https://www.sciencedirect.com/science/article/pii/0168900285906229}{A
  fast timing light pulser for scintillation detectors}, Nuclear Instruments
  and Methods in Physics Research Section A: Accelerators, Spectrometers,
  Detectors and Associated Equipment 241~(2) (1985) 612--613.
\newblock \href {https://doi.org/https://doi.org/10.1016/0168-9002(85)90622-9}
  {\path{doi:https://doi.org/10.1016/0168-9002(85)90622-9}}.
\newline\urlprefix\url{https://www.sciencedirect.com/science/article/pii/0168900285906229}

\bibitem{TIB}
L.~A. Tejedor, J.~A. Barrio, P.~Peñil, A.~Pérez, D.~Herranz, J.~Martín, A
  trigger interface board for the large and medium sized telescopes of the
  cherenkov telescope array, Nuclear Instruments and Methods in Physics
  Research Section A: Accelerators, Spectrometers, Detectors and Associated
  Equipment 1027 (2022) 166058.
\newblock \href {https://doi.org/https://doi.org/10.1016/j.nima.2021.166058}
  {\path{doi:https://doi.org/10.1016/j.nima.2021.166058}}.

\bibitem{xy_table}
B.~Biasuzzi, et~al., Design and characterization of a single photoelectron
  calibration system for the nectarcam camera of the medium-sized telescopes of
  the cherenkov telescope array, Nuclear Instruments and Methods in Physics
  Research Section A: Accelerators, Spectrometers, Detectors and Associated
  Equipment 950 (2020) 162949.
\newblock \href {https://doi.org/https://doi.org/10.1016/j.nima.2019.162949}
  {\path{doi:https://doi.org/10.1016/j.nima.2019.162949}}.

\bibitem{ticks}
C.~Champion, et~al., {TiCkS: A Flexible White-Rabbit Based Time-Stamping
  Board}, in: {16th International Conference on Accelerator and Large
  Experimental Physics Control Systems}, 2018, p. TUPHA090.
\newblock \href {https://doi.org/10.18429/JACoW-ICALEPCS2017-TUPHA090}
  {\path{doi:10.18429/JACoW-ICALEPCS2017-TUPHA090}}.

\bibitem{WhiteRabbit2011}
M.~Lipiński, et~al., White rabbit: a ptp application for robust sub-nanosecond
  synchronization, in: 2011 IEEE International Symposium on Precision Clock
  Synchronization for Measurement, Control and Communication, 2011, pp. 25--30.
\newblock \href {https://doi.org/10.1109/ISPCS.2011.6070148}
  {\path{doi:10.1109/ISPCS.2011.6070148}}.

\bibitem{PTP}
Ieee standard for a precision clock synchronization protocol for networked
  measurement and control systems, IEEE Std 1588-2019 (Revision ofIEEE Std
  1588-2008) (2020) 1--499\href {https://doi.org/10.1109/IEEESTD.2020.9120376}
  {\path{doi:10.1109/IEEESTD.2020.9120376}}.

\bibitem{JTAG}
Ieee standard for test access port and boundary-scan architecture - redline,
  IEEE Std 1149.1-2013 (Revision of IEEE Std 1149.1-2001) - Redline (2013)
  1--899.

\bibitem{daq}
D.~Hoffmann, et~al., 40 gbps data acquisition system for nectarcam, Journal of
  Physics: Conference Series 898~(3) (2017) 032015.
\newblock \href {https://doi.org/10.1088/1742-6596/898/3/032015}
  {\path{doi:10.1088/1742-6596/898/3/032015}}.

\bibitem{linearity}
F.~{Bradascio}, {Improved performances of the NectarCAM, a Medium-Sized
  Telescope Camera for the Cherenkov Telescope Array}, 2022, p.~25.
\newblock \href {https://doi.org/10.58027\%2Fx1th-cw93}
  {\path{doi:10.58027\%2Fx1th-cw93}}.

\bibitem{scipy}
P.~Virtanen, et~al., {SciPy} 1.0: fundamental algorithms for scientific
  computing in python, Nature Methods 17~(3) (2020) 261--272.
\newblock \href {https://doi.org/10.1038/s41592-019-0686-2}
  {\path{doi:10.1038/s41592-019-0686-2}}.

\bibitem{ICRC2019}
T.~Tavernier, et~al., {Status and performance results from NectarCAM -- a
  camera for CTA medium sized telescopes}, PoS ICRC2019 (2020) 805.
\newblock \href {https://doi.org/10.22323/1.358.0805}
  {\path{doi:10.22323/1.358.0805}}.

\bibitem{leo}
W.~R. Leo, {Techniques for nuclear and particle physics experiments: a how-to
  approach; 2nd ed.}, Springer, Berlin, 1994.
\newblock \href {https://doi.org/10.1007/978-3-642-57920-2}
  {\path{doi:10.1007/978-3-642-57920-2}}.

\bibitem{pmt_tt_simulations}
L.~A. Tejedor, et~al., {An Analog Delay Compensation System to Reduce the
  Effect of Variable Transit Time in PMTs}, IEEE Transactions on Nuclear
  Science 60~(4) (2013) 2905--2911.
\newblock \href {https://doi.org/10.1109/TNS.2013.2271300}
  {\path{doi:10.1109/TNS.2013.2271300}}.

\end{thebibliography}

\end{document}